%% file: chaos_Yang_Mills_v1.tex
\documentclass[12pt]{article}
\usepackage{graphicx}
\usepackage{placeins}
\usepackage{hyperref}
\usepackage{amsmath}
\usepackage{amssymb}
\usepackage{color}
\usepackage{physics}
\usepackage[nosort]{cite}
\usepackage{tikz}
\usetikzlibrary{arrows.meta,arrows,decorations.pathmorphing,decorations.pathreplacing,decorations.markings}
\tikzset{gl/.style={decorate,decoration={snake,amplitude=1.1pt,segment length=5pt}}}
\graphicspath{{numerics/}}

\newlength{\abstractwidth}
\newcommand{\raild}[4]{%
\foreach \y/\l/\m in {3/1/{#1}, 2.2/2/{#2}, 0.8/3/{#3}, 0/4/{#4}}
 {\draw[thick] (0,\y)--(7,\y); \node[left] at (0,\y) {\scriptsize \l};
  \node[right] at (7,\y) {\scriptsize $\m$};}}
\newcommand{\rungbox}[1]{\begin{tikzpicture}[xscale=1.02,yscale=.66,baseline=(current bounding box.center)]#1\end{tikzpicture}}
\newcommand{\raildmom}[4]{%
\foreach \y/\l/\m in {3/1/{#1}, 2.2/2/{#2}, 0.8/3/{#3}, 0/4/{#4}}
 {\draw[thick] (0,\y)--(7,\y); \node[left] at (0,\y) {\scriptsize $k_\l$};
  \node[right] at (7,\y) {\scriptsize $\m$};}}
\newcommand{\rungwide}[1]{\begin{tikzpicture}[xscale=1.08,yscale=.70,baseline=(current bounding box.center)]#1\end{tikzpicture}}
\newcommand{\dKone}{\draw[gl] (2.6,3)--(2.6,0.8);\node[left] at (2.55,1.9){\scriptsize$\ell_1$};
  \draw[gl] (4.4,2.2)--(4.4,0);\node[right] at (4.45,1.1){\scriptsize$\ell_2$};}
\newcommand{\dKonec}{\draw[gl] (2.6,3)--(4.6,0);\node[left] at (2.9,1.6){\scriptsize$\ell_1$};
  \draw[gl] (4.4,2.2)--(2.6,0.8);\node[right] at (3.9,1.4){\scriptsize$\ell_2$};}
\newcommand{\dKtwo}{\draw[gl] (2.4,3)--(2.4,0.8);\node[left] at (2.35,1.9){\scriptsize$\ell$};
  \draw[gl] (4.6,3)--(4.6,0.8);\node[right] at (4.65,1.9){\scriptsize$L{-}\ell$};
  \node at (3.5,3.32){\scriptsize$k_1{-}\ell$};\node at (3.5,0.48){\scriptsize$k_3{+}\ell$};}
\newcommand{\dKthree}{\draw[gl] (2.4,3)--(2.4,0.8);\node[left] at (2.35,1.9){\scriptsize$\ell_1$};
  \draw[gl] (4.6,3)--(4.6,0);\node[right] at (4.65,1.5){\scriptsize$\ell_2$};
  \node at (3.5,3.32){\scriptsize$k_1{-}\ell_1$};}
\newcommand{\dKfour}{\draw[gl] (3.6,3) to[bend right=42] (3.6,0.8);
  \draw[gl] (3.6,3) to[bend left=42] (3.6,0.8);
  \filldraw (3.6,3) circle (1.6pt);\filldraw (3.6,0.8) circle (1.6pt);
  \node[left] at (2.95,1.9){\scriptsize$\ell$};\node[right] at (4.25,1.9){\scriptsize$L{-}\ell$};}
\newcommand{\dKfivea}{\draw[gl] (3.3,3)--(2.7,0.8);\draw[gl] (3.6,3)--(4.6,0);
  \filldraw (3.45,3) circle (1.6pt);
  \node[left] at (2.8,1.9){\scriptsize$\ell_1$};\node[right] at (4.3,1.5){\scriptsize$\ell_2$};}
\newcommand{\dKfiveb}{\draw[gl] (3.2,3)--(2.6,0.8);\draw[gl] (3.5,3)--(4.4,0.8);
  \filldraw (3.35,3) circle (1.6pt);
  \node[left] at (2.7,1.9){\scriptsize$\ell$};\node[right] at (4.15,1.9){\scriptsize$L{-}\ell$};
  \node at (3.5,0.48){\scriptsize$k_3{+}\ell$};}
\newcommand{\dKbox}{\draw[thick] (0,3)--(7,3);\draw[thick] (0,0.8)--(7,0.8);
  \node[left] at (0,3){\scriptsize $A$};\node[left] at (0,0.8){\scriptsize $B$};
  \draw[gl] (3.5,3)--(3.5,0.8);\node[right] at (3.6,1.9){\scriptsize$\ell$};
  \node at (5.4,3.32){\scriptsize$q{-}\ell$};\node at (5.4,0.48){\scriptsize$-q{+}\ell$};}

\renewcommand{\tr}{\mathrm{Tr}}
\newcommand{\be}{\begin{equation}}
\newcommand{\ee}{\end{equation}}

\begin{document}

\begin{titlepage}
\begin{center}
{\Large\bf Chaos in Yang--Mills}\\[8mm]
Barel Skuratovsky\\[2mm]
{\it Department of Particle Physics and Astrophysics,\\
Weizmann Institute of Science, Rehovot 7610001, Israel}\\[3mm]
\texttt{bar-el.skuratovsky@weizmann.ac.il}\\[10mm]
\end{center}

\begin{center}
\begin{minipage}{\abstractwidth}
\noindent
We study the dynamics of pure-gauge $SU(N)$ through the time evolution of a spatial
Wilson loop. We compute its out-of-time-order correlator at weak 't~Hooft
coupling and large $N$, and propose a framework for the strong-coupling
description.  We find that operators in pure-gauge Yang--Mills grow only when the
group is non-abelian and $d>2$, and  conjecture that pure gauge theories
are chaotic only when non-abelian at $d>2$.  Even at weak coupling the correlator's
exponent inherits the IR sensitivity of the gluon self-energy imaginary
part, which rests on the non-perturbative magnetic scale. We then
explore the possibility that the loop's own size supplies that scale, since the loop operator is blind to any momenta softer than $1/2R$.  With
the cutoff imposed that way the exponent changes sign at a critical
diameter: smaller loops scramble, larger ones do not, and the scrambling
time diverges at the crossing. The transition follows from that cutoff.  Further investigation, in
particular a first-principles treatment of the thermal decay width, is
required to establish whether it is physical.
\end{minipage}
\end{center}

\vfill
\end{titlepage}

\tableofcontents
\newpage

\section{Introduction}
\label{sec:intro}

Out-of-time-order four-point functions diagnose chaos in many-body quantum
systems \cite{larkin_ovchinnikov_1969,shenker_stanford_2014,maldacena_bound_2016}.  The standard diagnostic is
the thermal expectation value of a squared commutator,
\begin{equation}
c(t)=\Big\langle\big[W(t),V\big]\big[W(t),V\big]^{\dagger}\Big\rangle_\beta\,,
\label{eq:cintro}
\end{equation}
which for simple operators in a large-$N$ system is expected to grow
exponentially over a long window,
$c(t)\propto N^{-2}e^{\lambda_Lt}$, before saturating at a value fixed by
factorisation; the same growth is visible in the size of the evolving
operator \cite{roberts_operator_2018}.  The rate $\lambda_L$ measures the strength of chaos, and is
conjectured to obey $\lambda_L\le2\pi/\beta$ \cite{maldacena_bound_2016}.
This work attempts to understand chaos in Yang--Mills by studying how operators and their OTOCs grow. In pure gauge, the simplest gauge-invariant operators are $F^{2}$ and closed Wilson lines. Wilson loops are the natural objects to discuss in large-$N$, as all gauge-invariant observables can be expressed in terms of these loops \cite{makeenko_methods_2002}. Moreover, they are order parameters for confinement \cite{wilson,aharony_hagedorndeconfinement_2004} and exhibit an area law in the IR.

OTOCs have been calculated in weakly coupled large-$N$ QFTs \cite{stanford_many-body_2016,chowdhury_swingle_2017,steinberg_thermalization_2019} by identifying the leading-order rung diagrams contributing to the four-point function and solving a Bethe--Salpeter equation (BSE) for the ladder they build.  The relation to the BFKL resummation is
taken up in \S\ref{sec:discussion}.  Holographically, the OTOC is instead read as a scattering amplitude of
extended objects, with the stringy
corrections of \cite{shenker_stringy_2015} (Regge behaviour, transverse
spreading) controlling the answer.  We discuss a string-like picture in
Sec.~\ref{sec:strong}.

Real-time gauge dynamics have been studied in different contexts before
\cite{berges_gaugeinvariant_2020}, and Wilson loops have been used as
chaos probes in holographic settings
\cite{hashimoto_chaoswilson_2018}.  Unlike a local field, the loop
carries its own scale, the size $R$ of the contour.  We therefore study the squared
commutator of a spatial Wilson loop with itself at two times,
\begin{equation}
c(t)=\frac1Z\tr\Big(e^{-\beta H/2}\big[W^t_\gamma,W^{0\dagger}_\gamma\big]
e^{-\beta H/2}\big[W^t_\gamma,W^{0\dagger}_\gamma\big]^{\dagger}\Big)\,,
\label{eq:cintroloop}
\end{equation}
in $SU(N)$ Yang--Mills at large $N$, weak 't~Hooft coupling
$\lambda=g^2N$, in $3+1$ dimensions at temperature $T$.

Evolving the loop in time displaces its contour, and
the term in that displacement that produces growth is a commutator of two
spatial components of the gauge field.  It therefore vanishes for an abelian
group, and in two spacetime dimensions there is only one spatial direction
for it to act on, so the growth stops, consistent with 2d Yang--Mills
having no propagating gluons at all \cite{witten_2dym_1991}.
We conclude that operator growth
in pure gauge theory requires a non-abelian group and $d>2$,
and we conjecture that pure gauge theories are chaotic only when
non-abelian at $d>2$; this is the content of Sec.~\ref{sec:dynamics}.

We follow the method of
\cite{stanford_many-body_2016}: resum the ladder of the four-point
function into a Bethe--Salpeter equation, put its free blocks on shell,
and read the Lyapunov exponent off the largest eigenvalue of the
resulting matrix; \cite{steinberg_thermalization_2019} applies it to a
gauge theory.  Expanding
the two loops to leading order turns
\eqref{eq:cintroloop} into a gluon four-point function whose eight fields
are contracted in $36$ surviving ways, falling into three topologies,
which we call classes (i), (ii) and (iii).  The
contracted lines running from time $t$ to time $0$ we call rails, and a
gluon exchanged between two of them a rung; a ladder is a stack of rungs
on a fixed set of rails.  Two
of the three topologies admit a ladder that can be resummed, and the
colour algebra decides how cheaply each ladder can be built.  In the
topology where each commutator contracts with itself, a single exchanged
gluon between the two commutators vanishes identically by the adjoint
trace, so the cheapest rung costs two gluons and the ladder starts at
$O(\lambda^2)$; in class (ii) a single dressed gluon is allowed
and the rung enters at $O(\lambda)$, the same order as the damping of
the lines it connects.  We solve the class (ii) equation numerically; the
class (i) equation is written down but not solved, for the reasons given
in \S\ref{sec:lyap}.  The colour counting runs the other way:
class (i) is the dominant of the two by colour, class (ii) being down by
$1/N^2$, so the exponent we report is not the leading one at large $N$;
the leading channel proved too complicated to calculate at present.

With the rails' damping
taken at its known value, the exponent is IR-sensitive even at weak
coupling, resting on the non-perturbative magnetic sector through the
gluon decay width and through the soft end of the integrals over the momentum
each rung carries.  We attempt to bypass that by imposing a physical
cutoff: the loop is a colour singlet of diameter $2R$, no exchange softer
than $1/2R$ couples to it, and when that scale stands in for the magnetic
one, the exponent changes sign at a critical loop diameter.  For class (ii),
the topology that allows the cheaper single-gluon rung, it is
\begin{equation}
2R^*m_D\;\simeq\;0.56\,\lambda^{-1/2}\,,
\label{eq:Rstarintro}
\end{equation}
with $m_D$ the Debye mass, whose inverse is the distance over which the
plasma screens an electric charge.  Here $m_D\propto
\sqrt\lambda\,T$ while the magnetic mass is $\propto\lambda T$, so
$m_D/m_{\rm mag}\propto\lambda^{-1/2}$.  Measured in Debye units, the
critical diameter therefore follows the magnetic screening length, not
the Debye length.  Loops smaller than $R^*$ scramble; larger loops decay at the
rate set by the damping of their constituent gluons.  The crossing is
continuous, so the scrambling time diverges as $R\to R^*$.

We will start by asking the seemingly simple question: how does a Wilson
loop evolve in time? 

The paper is organised as follows.  Section~\ref{sec:dynamics} derives the
displacement identity, isolates the term that produces growth, and
establishes that only non-abelian theories at $d>2$ grow.
Section~\ref{sec:weak} is the weak-coupling computation: the four-point
function, the surviving contractions and their poles, the rungs, the
Bethe--Salpeter equations, and the Lyapunov exponent.
Section~\ref{sec:strong} treats strong coupling, where gluons are not
the excitations, so it picks up the thread of Sec.~\ref{sec:dynamics}
rather than continuing Sec.~\ref{sec:weak}: it proposes the loop
two-point function as a cobordism sum and the OTOC as a $2\to2$ string
amplitude, and states the
obstruction, that the scattering sits above the cutoff of the effective
string description.  Sec.~\ref{sec:discussion} discusses the results and suggested future directions. 

\section{Dynamics and operator growth}
\label{sec:dynamics}

The observable of this paper is the spatial Wilson loop
\begin{equation}
    W_{\gamma}=\tr_{c}\,P\exp\Big(ig\oint_{\gamma}dx^{\mu}\,A^{a}_{\mu}T^{a}\Big),
    \label{loop}
\end{equation}
with the trace over colour and $P$ ordering the matrices along $\gamma$.  Two of its properties drive everything below.  It is non-local and carries a scale $R$, the contour radius, and it is gauge invariant. According to the Heisenberg equation, $\dot{W}_{\gamma}=i[H,W_{\gamma}]$, with the Yang--Mills Hamiltonian in $d$ spacetime dimensions
\begin{equation}
    H=\int d^{d-1}x\;\tr_{c}\big(E^{a}_{i}E^{a}_{i}+B^{a}_{i}B^{a}_{i}\big),
    \label{Hamiltonian}
\end{equation}

We evolve the fields and rebuild the loop out of them,
\begin{equation}
    W^{t}_{\gamma}=\tr_{c}\,P\exp\Big(ig\oint_{\gamma}dx^{\mu}\,e^{iHt}A^{a}_{\mu}(0)e^{-iHt}T^{a}\Big),
    \label{time_string}
\end{equation}
and show in Sec.~\ref{sec:heisen} that \eqref{time_string} equals the Heisenberg-evolved operator $e^{iHt}W_{\gamma}e^{-iHt}$ exactly, for a contour lying in a single time slice.  The contour is spacelike throughout; one with a timelike component carries fields at different times, and the identification no longer holds in this form.

Section~\ref{sec:twomodes} expands the exponent of \eqref{time_string} and finds two modes in it, a displacement of the contour and a source that cuts it.  Section~\ref{sec:heisen} identifies the first as the displacement operator and writes the evolved loop exactly. Section~\ref{sec:growth} shows that the second is what splits the loop in two, and that it is present only for non-abelian groups in $d>2$. Subsequently, we discuss the growth of the OTOC in Sec.~\ref{sec:weak}.

\subsection{Time evolution and the two modes}
\label{sec:twomodes}

Expanding the exponent of \eqref{time_string} in nested commutators,
\begin{align}
    ig\oint_{\gamma}dx^{\mu}\sum_{k=0}^{\infty}\frac{(it)^{k}}{k!}
    [\underbrace{H,[H,\dots[H}_{k},A^{a}_{\mu}(0)]\dots]]
    =\;&ig\oint_{\gamma}dx^{\mu}\Big(A^{a}_{\mu}(0)+it[H,A^{a}_{\mu}]
    \nonumber\\
    &-\frac{t^{2}}{2}[H,[H,A^{a}_{\mu}]]+O(t^{3})\Big),
    \label{exp1}
\end{align}
each commutator with $H$ contributes one power of $t$.  For the operator algebra we work in temporal gauge $A_{0}=0$, where the electric field is canonically conjugate to the connection,
\begin{equation}
    [A^{c}_{j}(\vec{x}),E^{b}_{k}(\vec{y})]=i\delta^{bc}\delta_{jk}\delta^{(d-1)}(\vec{x}-\vec{y}),
    \label{etcr}
\end{equation}
so that $E^{a}_{i}=F^{a}_{0i}=\dot{A}^{a}_{i}$, the equation of motion is $\dot{E}^{a}_{i}=(D_{j}F_{ji})^{a}$, and the Gauss law is the constraint $(D_{i}E_{i})^{a}=0$. We note that the diagrammatic computation of Sec.~\ref{sec:weak} is carried out in a covariant gauge instead; the observable is gauge invariant, and each gauge is used where it is convenient.  The first commutator is
\begin{equation}
    [H,A^{a}_{\mu}(0)]=-iE^{a}_{\mu}=-iF^{a}_{0\mu},
    \label{first_commute}
\end{equation}
and the second inserts the source,
\begin{equation}
    [H,[H,A^{a}_{\mu}]]=-i[H,E^{a}_{\mu}]=-(D_{j}F_{j\mu})^{a}=-\partial_{j}F^{a}_{j\mu}-gf^{abc}A^{b}_{j}F^{c}_{j\mu}.
    \label{second_commute}
\end{equation}
These represent the two displacement modes of the loop. The first moves the contour and changes nothing else about the operator; the second inserts field strength at a point of it that will be shown to cut the loop.  The two alternate along the series: odd orders load the electric field and even orders fire the source.

\subsection{The evolved loop and its displacement}
\label{sec:heisen}

Let us show that for a spatial contour $\gamma$, the loop obeys the Heisenberg equation
\begin{equation}
W_{\gamma}(t)\equiv e^{iHt}\,W_{\gamma}\,e^{-iHt}\;{=}\;\mathrm{tr}_{c}\,P\exp\Big(ig\oint_{\gamma}dx^{i}\,A_{i}(t,\vec{x})\Big).
\label{heisenloop}
\end{equation}
To do that we write the path-ordered exponential as the limit of ordered products $\prod_{k}\big(1+ig\,A(\vec x_{k})\!\cdot\!\Delta x_{k}\big)$ and insert $1=e^{-iHt}e^{iHt}$ between neighbouring factors. $H$ is a colour singlet and commutes with the fixed matrices $T^{a}$, with the ordering operator and with the trace. It therefore distributes onto each factor, $A_{i}(\vec x_{k})\mapsto A_{i}(t,\vec x_{k})$. The only property used is that every field in $W_{\gamma}$ sits on a single time slice, so one conjugation advances them all by the same $t$.

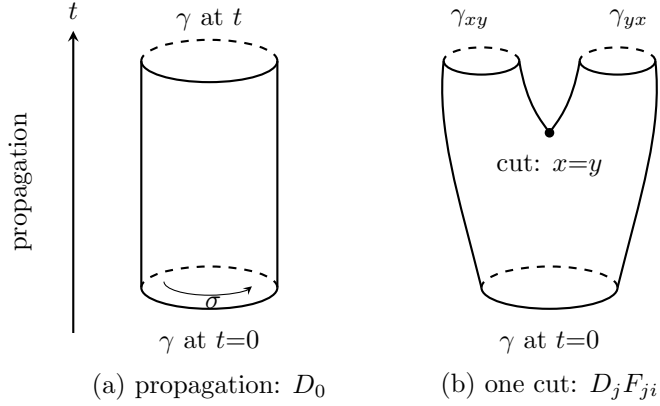
\begin{figure}[htbp]
\centering
\begin{tikzpicture}[scale=1.0,>=stealth,thick,every node/.append style={font=\footnotesize}]
  \draw[->] (-5.8,-0.6) -- (-5.8,3.4) node[above]{$t$};
  \node[rotate=90,anchor=south] at (-6.2,1.4) {propagation};

  \draw (-4.9,0) -- (-4.9,3);
  \draw (-3.1,0) -- (-3.1,3);
  \draw (-3.1,0) arc (0:-180:0.9 and 0.28);
  \draw[dashed] (-3.1,0) arc (0:180:0.9 and 0.28);
  \draw (-3.1,3) arc (0:-180:0.9 and 0.28);
  \draw[dashed] (-3.1,3) arc (0:180:0.9 and 0.28);
  \draw[->,thin] (-4.6,0.08) arc (180:340:0.6 and 0.18);
  \node at (-3.95,-0.2) {$\sigma$};
  \node at (-4,-0.72) {$\gamma$ at $t{=}0$};
  \node at (-4,3.55) {$\gamma$ at $t$};
  \node at (-4,-1.3) {(a) propagation: $D_{0}$};

  \draw (-0.4,0) .. controls (-0.7,1.2) and (-1.0,2.1) .. (-0.9,3);
  \draw (1.4,0) .. controls (1.7,1.2) and (2.0,2.1) .. (1.9,3);
  \draw (0.1,3) .. controls (0.2,2.4) and (0.4,2.2) .. (0.5,2.05);
  \draw (0.9,3) .. controls (0.8,2.4) and (0.6,2.2) .. (0.5,2.05);
  \draw (1.4,0) arc (0:-180:0.9 and 0.28);
  \draw[dashed] (1.4,0) arc (0:180:0.9 and 0.28);
  \draw (0.1,3) arc (0:-180:0.5 and 0.18);
  \draw[dashed] (0.1,3) arc (0:180:0.5 and 0.18);
  \draw (1.9,3) arc (0:-180:0.5 and 0.18);
  \draw[dashed] (1.9,3) arc (0:180:0.5 and 0.18);
  \filldraw (0.5,2.05) circle (1.4pt);
  \node at (0.5,1.62) {cut: $x{=}y$};
  \node at (0.5,-0.72) {$\gamma$ at $t{=}0$};
  \node at (-0.55,3.55) {$\gamma_{xy}$};
  \node at (1.55,3.55) {$\gamma_{yx}$};
  \node at (0.5,-1.3) {(b) one cut: $D_{j}F_{ji}$};
\end{tikzpicture}
\caption{Worldsheet of the time-evolved Wilson loop, drawn as a string splitting amplitude. (a) Pure propagation: the loop $\gamma$ is rigidly displaced along the time direction $t$, sweeping a cylinder, the displacement mode $D_{0}=ig\,F_{0i}\dot{x}^{i}$ of Eq.~\eqref{displacement_op}. (b) One cutting event: where two points of the loop coincide in space ($x{=}y$, the $\delta^{(d)}(x-y)$ of the Makeenko--Migdal equation) the loop pinches and splits, $\gamma\to\gamma_{xy}\cup\gamma_{yx}$, and the tube becomes a pair of pants, the source mode $D_{j}F_{ji}\subset\nabla_{\mu}F_{\mu\nu}$ of Eq.~\eqref{loop_space_EOM}. The worldsheet coordinates are the loop parameter $\sigma$ (around each boundary) and the propagation time.}
\label{fig:pants}
\end{figure}

The leading term has geometric meaning,  under a deformation of the contour, $x^{\mu}(\tau)\to x^{\mu}(\tau)+\delta x^{\mu}(\tau)$, varying the holonomy and integrating by parts inside the path ordering gives
\begin{equation}
    \delta_{\gamma}W[\gamma]=ig\oint_{\gamma}d\tau\,\delta x^{\nu}(\tau)\,\Big[P\,F_{\nu\mu}\big(x(\tau)\big)\,\dot{x}^{\mu}(\tau)\,\cdots\Big]
    \equiv \oint_{\gamma}d\tau\,\delta x^{\nu}(\tau)\,D_{\nu}(\tau),
    \label{contour_variation}
\end{equation}
which defines the displacement operator of the line,
\begin{equation}
    D_{\nu}(\tau)=ig\,F_{\nu\mu}\big(x(\tau)\big)\,\dot{x}^{\mu}(\tau).
    \label{displacement_op}
\end{equation}
Only the transverse part of $\delta x^{\nu}$ contributes: a longitudinal $\delta x^{\nu}\propto\dot{x}^{\nu}$ gives $F_{\nu\mu}\dot{x}^{\nu}\dot{x}^{\mu}=0$, killed by reparametrisation invariance to all orders. The leading order term in $t$ of \eqref{exp1} is precisely this operator with $\nu=0$. For a static spacelike loop ($\dot{x}^{0}=0$),
\begin{equation}
    ig\oint_{\gamma}dx^{i}\,t\,F^{a}_{0i}=t\oint_{\gamma}d\tau\,D_{0}(\tau),\qquad D_{0}=ig\,F_{0i}\dot{x}^{i},
    \label{first_order_displacement}
\end{equation}
So the loop is carried rigidly to the time-$t$ slice. We see that we can write the displacement built from the evolved field and integrated along the time translation,
\begin{equation}
    W^{t}_{\gamma}=\tr_{c}\,P\exp\left(ig\oint_{\gamma}A^{a}_{\mu}(0)\,dx^{\mu}+\oint_{\gamma}d\tau\int_{0}^{t}dt'\,D_{0}(\tau,t')\right),
    \label{displacement_exponentiated}
\end{equation}
with $D_{0}(\tau,t')=ig\,F_{0i}(t',x(\tau))\,\dot{x}^{i}$. To tie back to rigid displacement and splitting, we will examine at $O(t^{2})$ the source $D_{j}F_{ji}$. It is the spatial part of $\nabla_{\mu}F_{\mu\nu}$, and it inserts field strength at a point of the contour; by the Mandelstam formula and the path derivative \cite{makeenko_methods_2002} it is equivalent to inserting an area derivative, the loop-space operator
\begin{equation}
      \frac{i}{N}\,\tr\,P\,\nabla_{\mu}F_{\mu\nu}(x)\,e^{ig\oint A}=\partial^{x}_{\mu}\frac{\delta}{\delta\sigma_{\mu\nu}(x)}\,W_{\gamma} 
      \label{loop_space_EOM}
\end{equation}
whose average obeys the Makeenko--Migdal equation and therefore sources cuts: the expectation value satisfies 
\begin{equation}
   \partial^{x}_{\mu}\frac{\delta}{\delta\sigma_{\mu\nu}(x)}\,\big\langle W_{\gamma}\big\rangle = \lambda\oint_{\gamma}dy_{\nu}\,\delta^{d}(x-y)\,\big\langle W_{\gamma_{xy}}W_{\gamma_{yx}}\big\rangle
\end{equation}
To recap, the first mode is the first derivative of the connection, the displacement \eqref{first_order_displacement}; the second is the first evolved by one more step in time, the source that cuts the loop through the Makeenko--Migdal equation in the $N\rightarrow\infty$ limit. If you think about the spatial loop as a string,  Figure~\ref{fig:pants} would be that string's worldsheet picture: pure propagation sweeps a cylinder, one cutting event turns the cylinder into a pair of pants, as in a string splitting amplitude.

\subsection{The loop splits only when non-abelian and $d>2$}
\label{sec:growth}

Throughout we work at large $N$, $N\to\infty$ with $\lambda=g^{2}N$ fixed. A cut produces two definite loops $\gamma_{xy},\gamma_{yx}$ through factorisation. To discuss whether the loop operator grows, we expand the loop itself in nested commutators, as \S\ref{sec:twomodes} expanded its exponent,
\begin{equation}
W^t_\gamma=\sum_{k\ge0}\frac{(it)^k}{k!}\,\mathrm{ad}_H^k\,W_\gamma\,,
\qquad \mathrm{ad}_H X\equiv[H,X]\,,
\label{eq:adexpansion}
\end{equation}
each commutator with $H$ does one of two things to the operator.  It can
displace the contour, through the electric term $E_i=F_{0i}$ of
\eqref{displacement_op}, which moves the loop without changing what it is; or it can insert the magnetic source
\begin{equation}
\dot E_i \;=\; D_jF_{ji}\;=\;\partial_jF_{ji}+ig\big[A_j,F_{ji}\big]\,,
\label{eq:source}
\end{equation}
which adds field strength at a point of the contour.  The first two
commutators make this explicit:
\begin{align}
i[H,W_\gamma]&=ig\oint_\gamma dx^i\;\tr\,P\Big[F_{0i}(x)\,e^{ig\oint A}\Big]\,,
\label{eq:HW}\\
i[H,[H,W_\gamma]]&=(ig)^2\oint_\gamma dx^i\!\oint_\gamma dy^k\;
\tr\,P\Big[F_{0i}(x)\,F_{0k}(y)\,e^{ig\oint A}\Big]
+ig\oint_\gamma dx^i\;\tr\,P\Big[(D_jF_{ji})(x)\,e^{ig\oint A}\Big]\,,
\label{eq:HHW}
\end{align}
with $P$ ordering each insertion into the exponential at its own contour
point: the first is a single displacement insertion, (a) of
Fig.~\ref{fig:pants}, and the second contains the displacement applied
twice and one insertion of \eqref{eq:source}, (b) of
Fig.~\ref{fig:pants}. We observe that the loop operator grows only when the second term is present, When the deformation
driven by \eqref{eq:source} brings two arcs of the contour into contact, the
colour indices at the contact point are exchanged by the Fierz identity for
$SU(N)$,
\begin{equation}
T^a_{ij}T^a_{kl}=\tfrac12\,\delta_{il}\delta_{kj}
-\tfrac1{2N}\,\delta_{ij}\delta_{kl}\,,
\label{eq:fierz}
\end{equation}
whose first term replaces one trace by the product of two,
$\mathrm{tr}\to\tfrac12\,\mathrm{tr}\,\mathrm{tr}$, while the second returns
the original loop suppressed by $1/N^2$.  A contact is therefore a
cut: the operator becomes two loops.  Repeating the process produces
the branching structure of Fig.~\ref{fig:worldsheet}, and the counting is
fixed by \eqref{eq:adexpansion}: each cut costs two powers of
$\mathrm{ad}_H$, hence $t^{2}$ and, once its colour cycle is closed, one
power of $\lambda$, so a history with $c$ cuts enters at order
$t^{2c}\lambda^{c}$.  (This identifies the channel that splits fastest)

The same counting can be read off the fields themselves, each source event
replaces one field by three, and each intervening derivative only converts
an $A$ into an $E$ to arm the next one, so the events are spaced two
commutators apart. To show how the number of gauge fields building up the time evolved connection increases, we follow 
$\partial_{t}^{2}(D_{j}F_{j\mu})$ along its commutations,
\begin{equation}
    g^{2}[A_{j},[A_{j},A_{\mu}]]
    \;\xrightarrow{\;\partial_{t}\;}\;
    g^{2}[E_{j},[A_{j},A_{\mu}]]
    \;\xrightarrow{\;\partial_{t}\;}\;
    g^{2}\big[(D_{k}F_{kj}),[A_{j},A_{\mu}]\big]
    \;\supset\;
    g^{4}\big[[A_{k},[A_{k},A_{j}]],[A_{j},A_{\mu}]\big],
    \hspace{-6mm}
    \label{fourth_expanded}
\end{equation}
a five-field word at order $g^{4}$: the maximal content at order $k$ is
$2\lfloor k/2\rfloor+1$ fields. 
\input{worked/worldsheet_figure.tex}

For an abelian group the structure constants vanish,
$[A_j,F_{ji}]=0$, and \eqref{eq:source} reduces to
$\partial_jF_{ji}$, a c-number. Meaning that for the abelian loop the time evolution is by a constant phase, $W_\gamma=\exp(ig\oint A)$, so there are no colour
indices at a contact for \eqref{eq:fierz} to exchange. Abelian gauge theory does not
split, and hence operators do not grow in any dimension. Throughout, $d$ counts spacetime dimensions, so
$d=2$ is the theory with a single spatial direction.  The source of
\eqref{eq:source} requires two distinct spatial indices, $F_{ji}$ with
$j\neq i$; with one spatial direction there is no magnetic field at all,
meaning $\dot E_i=0$ on the contour, and the second
commutator contains no source. Evolution is then pure displacement: the loop moves and its
area changes, but it never acquires the field strength that a contact
requires. It makes sense, as the two-dimensional theory is solvable, and the
statement can be made exactly. In 
$d=2$ the only field strength is $F_{01}$, the Hamiltonian is purely electric,
$H=\tfrac{1}{2}\int\!dx\,E^2$, and the electric field is conserved,
$\dot E=i[H,E]=0$; the source term of \eqref{eq:source} vanishes
identically and the loop can only 'slide'. Its evolution is then a pure
displacement of the contour, so the correlator depends on time only through
the swept area,
$\langle W^t_\gamma W^0_\gamma\rangle=\sum_{R}(\dim R)\,
e^{-\frac{g^2}{2}C_2(R)A(t)}$ \cite{witten_2dym_1991}.  It does not scramble:
the branching of
Fig.~\ref{fig:worldsheet} never starts. In $d>2$ a second spatial
direction exists, $\dot E_i=D_jF_{ji}\neq0$, and the source switches on. Now we can make a simple statement related to both the symmetry group and the dimension. In pure gauge theory operators grow only when the symmetry group is non-abelian and $d>2$.
\begin{equation}
\text{loop operators split}\quad\Longleftrightarrow\quad
\text{non-abelian}\ \ \text{and}\ \ d>2\,.
\label{eq:twoswitches}
\end{equation}
The situation is completely different when matter is present, then the contact can be resolved by pair creation instead of by \eqref{eq:fierz}, which enables splitting even in the abelian theory.  That case is outside the
scope of this paper and is commented on in Sec.~\ref{sec:discussion}.

What \eqref{eq:twoswitches} does not say is how fast these operators scramble. In the following Sec.~\ref{sec:weak} we work our way to computing the Lyapunov exponent in the weak coupling regime in 3+1. We see the exponent is IR sensitive even at weak coupling, and that if we take the loop size as an IR regulator it changes
sign at a critical loop size set by the plasma's screening lengths. 

\section{Weak coupling}
\label{sec:weak}

At weak coupling the loop operators are built from gluons and the
out-of-time-order correlator can be computed diagrammatically.  The
logic of this section is the standard one for Lyapunov exponents in
thermal field theory: identify the contractions whose ladders build up with
time, construct the rungs those ladders are built from, resum them into a
Bethe--Salpeter equation, and read the exponent off the BSE eigenvalues.  Two features are specific to loop operators and
drive everything below: the operators are non-local, so a loop size
$R$ enters as a second scale beside the temperature; and they are colour
singlets built from two adjoint insertions, so the colour algebra decides which rungs exist at all.  We note that even though we described the loop splitting at the previous section, we assume the contour stays fixed at least up to the scrambling time. To try and understand if this is reasonable, we will discuss the proposal to treat loop correlators as cobordisms in
\S\ref{sec:cobordism}. 

\subsection{The four-point function}
\label{sec:4pt}

We work in pure $SU(N)$ Yang--Mills in $3+1$ dimensions at large $N$,
keeping the diagrams that are leading in powers of $N$. The temperature is $T=1/\beta$ (units $T=1$ below), and the 't~Hooft 
coupling is taken to be small $\lambda=g^2N\ll1$. The diagrammatic computation of this section
is performed in a general linear covariant gauge, with the free propagator
parametrised as $P^T+P^L+\xi\,\ell\ell/(\ell^2)^2$, and we set $\xi=0$ throughout. We note this is a different gauge from the
temporal one used for the Hamiltonian analysis of
Sec.~\ref{sec:dynamics}: there the object of interest was the operator
algebra, where $A_0=0$ makes the electric field canonical, whereas here it
is the propagator, where the covariant form makes the transverse and
longitudinal structures explicit and the Landau choice is adopted for simplicity.  Nothing depends on either choice: the observable defined below is gauge
invariant. The contour $\gamma$ is a fixed
spatial curve $x^i(s)$ and we choose it to be a circle of radius $R$. We work in the following conventions
\begin{equation}
    \tr(T^aT^b)=\tfrac12\delta^{ab}, \tr T^a=0,
\int_q\equiv\int\!\frac{d^3q}{(2\pi)^3}, n_B(\omega)=(e^{\beta\omega}-1)^{-1},
P^T_{ij}(\hat q)=\delta_{ij}-\hat q_i\hat q_j
\end{equation}
The chaos diagnostic is the thermal square of the commutator of the loop
with itself at two times, with the thermal weight split symmetrically
across the two commutators,
\begin{equation}
c(t)\;=\;\frac{1}{Z}\,\tr\Big(e^{-\beta H/2}\,\big[W^t_\gamma,W^{0\dagger}_\gamma\big]\,
e^{-\beta H/2}\,\big[W^t_\gamma,W^{0\dagger}_\gamma\big]^{\dagger}\Big).
\label{eq:cdef}
\end{equation}
Expanding each loop to leading non-trivial order is what makes
\eqref{eq:cdef} a gluon four-point function.  The only input this section
takes from Sec.~\ref{sec:dynamics} is \eqref{heisenloop}: the loop at time
$t$ is the loop of the evolved connection, so the fields below are
Heisenberg fields carried to the time-$t$ slice, and no separate evolution
of the contour is needed. The linear term of
the path-ordered exponential vanishes identically
($\oint ds\,\tr T^a=0$), so each trace starts at two fields,
\begin{equation}
W^t_\gamma=1-\frac{\lambda}{2N^2}\oint\!\!\oint ds_1ds_2\;
\dot x^i(s_1)\dot x^j(s_2)\,A^a_i(x(s_1),t)A^a_j(x(s_2),t)+O(g^3),
\label{eq:Wexp}
\end{equation}
where $\oint\!\!\oint$ here runs over the ordered region $s_1>s_2$, and
likewise for $W^{0\dagger}_\gamma$ with the opposite ordering.  
Each loop lies on a single time slice, and the gauge field commutes with
itself at equal times, $\big[A^a_i(\vec x),A^b_j(\vec y)\big]=0$, so the
integrand above is symmetric under exchanging $s_1$ and $s_2$ and the
ordered integral is exactly half the unrestricted one.  Below we
therefore let $\oint\!\!\oint$ run unrestricted, at the cost of one
factor $\tfrac12$ per loop.  Substituting \eqref{eq:Wexp} into
\eqref{eq:cdef} gives the object this section computes,
\begin{multline}
c(t)=\frac{\lambda^4}{256N^8}\oint\!\!\oint ds_1ds_2\,\dot x_1^{\mu}\dot x_2^{\nu}
\oint\!\!\oint ds_{1'}ds_{2'}\,\dot x_{1'}^{\rho}\dot x_{2'}^{\sigma}
\oint\!\!\oint ds_3ds_4\,\dot x_3^{\alpha}\dot x_4^{\beta}
\oint\!\!\oint ds_{3'}ds_{4'}\,\dot x_{3'}^{\gamma}\dot x_{4'}^{\delta}\\[2pt]
\times\frac{1}{Z}\tr\Big(e^{-\beta H/2}
\big[A^{a}_{\mu}(t,x(s_{1}))A_{a\nu}(t,x(s_{2})),\,
A^{b}_{\rho}(0,x(s_{1'}))A_{b\sigma}(0,x(s_{2'}))\big]\\[2pt]
\times e^{-\beta H/2}
\big[A^{c}_{\alpha}(t,x(s_{3}))A_{c\beta}(t,x(s_{4})),\,
A^{d}_{\gamma}(0,x(s_{3'}))A_{d\delta}(0,x(s_{4'}))\big]^{\dagger}\Big).
\label{eq:master}
\end{multline}
 In most of the following we compute the eight-index thermal correlator (second line of \eqref{eq:master}), and its exponential growth in $t$. We note that in a covariant gauge the Lagrangian also carries Faddeev--Popov ghosts,
and they play no part below.  The loop operator is built from $A$ alone, so
no rail (gluon lines that carry the rungs in ladder diagrams) is a ghost; and a ghost cannot be exchanged between two rails
either, since ghost lines close on themselves and cannot begin or end on a
gluon line.  The only place one can appear at the order worked here is inside the dressing of a gluon line, where it is already contained in the self-energy. Turning to the interactions, from
$\mathcal L=-\tfrac14F^a_{\mu\nu}F^{a\mu\nu}$ with
$F^a_{\mu\nu}=\partial_\mu A^a_\nu-\partial_\nu A^a_\mu
+gf^{abc}A^b_\mu A^c_\nu$, the cubic and quartic interactions are
$\mathcal L_3=-gf^{abc}(\partial_\mu A^a_\nu)A^{b\mu}A^{c\nu}$ and
$\mathcal L_4=-\tfrac14g^2f^{abe}f^{cde}A^a_\mu A^b_\nu A^{c\mu}A^{d\nu}$,
giving the momentum-space vertex tensors (all momenta incoming,
$p+q+r=0$ and $p+q+r+w=0$)
\begin{align}
\Gamma_3^{abc,\mu\nu\rho}(p,q,r)&=-gf^{abc}\Big[
g^{\mu\nu}(p-q)^\rho+g^{\nu\rho}(q-r)^\mu+g^{\rho\mu}(r-p)^\nu\Big],
\label{eq:G3}\\
\Gamma_4^{abcd,\mu\nu\rho\sigma}&=-g^2\Big[
f^{abe}f^{cde}\big(g^{\mu\rho}g^{\nu\sigma}-g^{\mu\sigma}g^{\nu\rho}\big)
+f^{ace}f^{bde}\big(g^{\mu\nu}g^{\rho\sigma}-g^{\mu\sigma}g^{\nu\rho}\big)
\nonumber\\&\hspace{12mm}
+f^{ade}f^{bce}\big(g^{\mu\nu}g^{\rho\sigma}-g^{\mu\rho}g^{\nu\sigma}\big)\Big].
\label{eq:G4}
\end{align}
The metric is $g_{\mu\nu}=\mathrm{diag}(+1,-1,-1,-1)$ throughout, and
\eqref{eq:G3}--\eqref{eq:G4} are the standard $SU(N)$ Yang--Mills vertices
in that signature \cite{makeenko_methods_2002}, written with all momenta
incoming, with two differences of convention: We define these tensors real: every factor of $i$ generated by the
expansion of the time-evolution operators is collected into a single phase
per insertion, $\Phi_{n_v}$. 

\subsection{Surviving contractions: the rails}
\label{sec:rails}

\subsubsection{The three classes, and which of them we keep}
\label{sec:classes}

At leading order the eight fields of \eqref{eq:master} are contracted
pairwise.  Of the $105$ pairings, $69$ cancel between the terms of the two
commutators, and $36$ survive; they fall into three topologies:
\begin{center}
\begin{tabular}{llcl}
\hline
class & topology & count & long lines ($t\to0$)\\
\hline
(i) & each commutator contracts with itself & $4$ & four\\
(ii) & one long line per commutator, two caps & $16$ & two\\
(iii) & crossed ladder between the commutators & $16$ & --\\
\hline
\end{tabular}
\end{center}
\input{worked/contour_figure.tex}

We call a contraction joining the time-$t$ loop to the time-$0$ loop a rail: rails
are the lines that can carry the growth, because their length in time
grows with $t$.  Class (i) has four
rails (two per commutator); class (ii) has two rails joined by two
equal-time caps.  Class (iii) is not treated in this work: its
crossed topology does not present a fixed rail skeleton on which a ladder
can be iterated, and it is out of scope for this paper. The colour cycle of each class is closed by the singlet traces of
\eqref{eq:master}. The vanishing trace of a single adjoint generator and a
per-placement evaluation of the colour cycle are important:
\begin{itemize}
\item On class (i) a single gluon exchanged between the two
commutators carries $\tr_{\rm adj}F^e=0$, the trace of the adjoint
generator $(F^e)_{ab}=f^{aeb}$, and vanishes identically, to all
orders in $g$.  The leading rung therefore exchanges two
gluons and costs $g^4$; with the colour factor $N^2$ this is $O(\lambda^2)$.
\item On class (ii) a single exchanged gluon is allowed: class (ii)'s colour
cycle closes through the caps and gives
$-N\delta^{e_1e_2}$, i.e.\ $-\lambda$ per insertion.  The leading order rung is $O(\lambda)$.
\end{itemize}
Since the rails are damped at $O(\lambda)$ by soft scattering off the bath,
class (ii) is the channel in which gain and loss enter at the same
order, while class (i) starts one power of $\lambda$ behind.  On this ground
alone class (ii) is expected to grow faster, and the numerical results of
\S\ref{sec:lyap} bear that out.  By colour the prefactor of class (ii) is
down by $1/N^2$ relative to class (i); we keep it because its exponent
enters one order lower in $\lambda$.

\subsubsection{The free block: four rails on shell}
\label{sec:free}

Fourier transform each rail in space and Laplace transform in time,
$k_j=(k_j^0,\vec q_j)$, with $\omega=\sum_jk_j^0$ conjugate to $t$.  The time
direction is Laplace transformed because we expect the four-point function to be a
growing exponential which has no Fourier transform. Every rail attaches to the contour through the loop current
$J^i(\vec q\,)=\oint ds\,\dot x^i(s)\,e^{i\vec q\cdot x(s)}$, evaluated for a
circle in \eqref{eq:Jcurrent} below, and that current is conserved:
\begin{equation}
q_iJ^i(\vec q\,)=\oint\!ds\,\dot x^i(s)\,q_i\,e^{i\vec q\cdot x(s)}
=-i\oint\!ds\,\frac{d}{ds}e^{i\vec q\cdot x(s)}=0\,,
\label{eq:ward}
\end{equation}
the integrand being a total derivative around a closed curve.  With
$P^L_{ij}=\hat q_i\hat q_j$ this gives $J\,P^L\,J=(\hat q\cdot J)^2=0$, and
the same argument removes the gauge term $\xi\,q_iq_j/(q^2)^2$.  The rails are
transverse, and each rail carries $P^T$ at its
own momentum; we also solve for the scalar coefficient of the product of the four
projectors.

Which line is which follows from the operator ordering.  In each commutator
one line is retarded and the other carries the symmetric thermal weight,
\begin{equation}
F_0^{(i)}(k_1,k_2,k_3,k_4)=\Delta^R_{ij}(k_1)\,\widetilde D_{kl}(k_2)\,
\Delta^R_{mn}(k_3)\,\widetilde D_{pq}(k_4)\,,
\label{eq:F0i}
\end{equation}
\begin{equation}
\Delta^R_{ij}(k)=P^T_{ij}(\vec q\,)\,\Delta^R_T(k)\,,
\qquad
\widetilde D_{ij}(k)=P^T_{ij}(\vec q\,)\Big[n_B(k^0)+\tfrac12\Big]\rho_T(k)\,,
\label{eq:raillines}
\end{equation}
\begin{equation}
\Delta^R_T(k)=\frac{1}{q^2+\Pi_T(k^0,q)-(k^0+i0^+)^2}\,,
\qquad
\rho_T(k)=2\,\mathrm{Im}\,\Delta^R_T(k)\,,
\label{eq:DRTexplicit}
\end{equation}
with $\Pi_T$ the transverse hard-thermal-loop self-energy
\cite{braaten_soft_1990,lebellac_thermal_1996,blaizot_iancu_2002}.  The occupancy is $n_B+\tfrac12$ because
expanding a commutator gives both orderings, so the partner line is the
symmetric combination
\begin{equation}
\tfrac12\big[G^>(k)+G^<(k)\big]=\Big[n_B(k^0)+\tfrac12\Big]\rho_T(k)\,.
\label{eq:symmetric}
\end{equation}
Class (ii) has two rails and two equal-time contractions, drawn at the ends of
Fig.~\ref{fig:contour}.  Its two rails are the retarded pair; the
contractions are the thermal lines and carry no frequency, since each has
already been integrated over its own,
\begin{equation}
F_0^{(ii)}=\Delta^R_T(k_A)\,\Delta^R_T(k_B)\,B(q)^2\,P^TP^T\,,
\qquad
B(q)=\int\!\frac{d\ell^0}{2\pi}\,\frac{\rho_T(\ell^0,q)}{2\sinh(\beta\ell^0/2)}\,.
\label{eq:F0box}
\end{equation}
Near its pole the dressed retarded propagator is
\begin{equation}
\Delta^R_T(k)\simeq\frac{Z_T(q)}{\omega_T(q)^2-(k^0+i\gamma(q))^2}
=\frac{Z_T}{2\omega_T}\sum_{s=\pm}\frac{(-s)}{k^0-s\,\omega_T+i\gamma(q)}\,,
\label{eq:pfrac}
\end{equation}
with $\gamma(q)$ the damping rate \eqref{eq:gammafull} below, $Z_T$ the
residue, and $1/2\omega_T$ from writing the quadratic pole as two simple
ones.

Each rail joins a field at time $t$ to a field at time $0$, and we use
the propagators in frequency space, where the HTL functions are known:
every rail brings one $\int\dd k_j^0/2\pi\;e^{-ik_j^0t}$.  The product
depends on $t$ only through $e^{-i(\sum_jk_j^0)t}$, so the Laplace
transform fixes the sum, $\sum_jk_j^0=\omega$, and the three relative
frequencies still have to be integrated.  Class (i) has four rail
frequencies and one constraint, so three are free. Two of them are fixed by the thermal rails, which are on-shell delta functions: $\tfrac12\coth(\beta k^0/2)$ and $\rho_T$
are both odd, so their product is even, and
\begin{equation}
\widetilde D_T(k)\simeq 2\pi Z_T(q)\Big[n_B(\omega_T(q))+\tfrac12\Big]
\delta\big((k^0)^2-\omega_T^2(q)\big)\,.
\label{eq:deltarail}
\end{equation}
The third is fixed by the retarded pair.  Rail~1's poles lie $\gamma(q_1)$
below the real axis and rail~3's lie $\gamma(q_3)$ above, so the contour is
caught between them, as in Fig.~\ref{fig:pinch}, and can be approximated by delta
function,
\begin{multline}
\Delta^R_T(k_1^0;q_1)\,\Delta^R_T(\Omega-k_1^0;q_3)\;\simeq\;
-2\pi i\sum_{s_1,s_3=\pm}s_1s_3\,
\frac{Z_T(q_1)}{2\omega_T(q_1)}\frac{Z_T(q_3)}{2\omega_T(q_3)}\\
\times\;\frac{\delta\big(k_1^0-s_1\omega_T(q_1)\big)}
{\omega-k_2^0-k_4^0-s_1\omega_T(q_1)-s_3\omega_T(q_3)+i\big[\gamma(q_1)+\gamma(q_3)\big]\;}\,,
\label{eq:R13}
\end{multline}
 The replacement keeps the part that governs the correlator at large
time.

\input{worked/pinch_contour_figure.tex}

The denominator reaches its pole when the four frequencies cancel,
\begin{equation}
\sum_j s_j\,\omega_T(q_j)=0\,.
\label{eq:pinch}
\end{equation}
Each rail sits at $k_j^0=s_j\,\omega_T(q_j)$, and the denominator of
\eqref{eq:R13} is $\omega-\sum_js_j\omega_T(q_j)$ plus the widths, so the
assignments obeying \eqref{eq:pinch} are the ones that survive at large
time.  At total momentum near zero the two rails of each contraction are
back to back,
\begin{equation}
\vec q_2=-\vec q_1\,,\qquad \vec q_4=-\vec q_3\,,
\label{eq:backtoback}
\end{equation}
The retarded rails
carry the sign weight $s_1s_3$ of \eqref{eq:R13} and the thermal rails
carry none, and adding $s_1s_3$ over the assignments that obey
\eqref{eq:pinch} at $\omega_T(q_1)=\omega_T(q_3)$ gives $-2$.  Collecting
\eqref{eq:deltarail} and \eqref{eq:R13}, the free block is a product of on-shell delta functions over a single pole,
\begin{multline}
F_0^{(i)}\;\simeq\;-2\pi i\sum_{s_1,s_3=\pm}s_1s_3\,
\frac{Z_T(q_1)}{2\omega_T(q_1)}\frac{Z_T(q_3)}{2\omega_T(q_3)}\,
\delta\big(k_1^0-s_1\omega_T(q_1)\big)\\
\times\prod_{j=2,4}2\pi Z_T(q_j)\Big[n_B(\omega_T(q_j))+\tfrac12\Big]
\delta\big((k_j^0)^2-\omega_T^2(q_j)\big)\;
\frac{1}{\;\omega+i\sum_j\gamma(q_j)\;}\,.
\label{eq:F0onshell}
\end{multline}
The width sums over all four rails.  Rails 1 and 3 bring
$\gamma(q_1)+\gamma(q_3)$ through \eqref{eq:R13}.  Rails 2 and 4 are the
same narrow Lorentzians, \eqref{eq:pfrac}, and \eqref{eq:deltarail} keeps
only their area; treating them as strict deltas would lose their widths,
so we keep $\gamma(q_2)+\gamma(q_4)$ as well, and with
\eqref{eq:backtoback} the total is
$\sum_j\gamma(q_j)=2\gamma(q_1)+2\gamma(q_3)$.  The width is not part of
the HTL dressing, whose self-energy is real at the timelike pole:
$\gamma$ is the rate at which the plasma scatters the rail gluon, one
soft exchange, taken from the known calculation, \eqref{eq:gammafull}.

Class (ii) is the same calculation with one free frequency instead of three.
Its rails carry $\pm\vec q$, so both sit at the same $\omega_T(q)$ with no
further condition; the surviving sector is $s_B=-s_A$, and
\begin{equation}
F_0^{(ii)}\;\simeq\;2\pi i\,B(q)^2
\Big[\frac{Z_T(q)}{2\omega_T(q)}\Big]^2
\frac{\delta\big(k_A^0-\omega_T(q)\big)+\delta\big(k_A^0+\omega_T(q)\big)}
{\;\omega+2i\gamma(q)\;}\,.
\label{eq:F0boxonshell}
\end{equation}
Analogous to eq.~(20) of \cite{stanford_many-body_2016}, with $Z_T/2\omega_T$
in place of $1/2E$; the overall sign differs because the retarded propagator
there carries a factor of $i$ relative to \eqref{eq:pfrac}. The on shell delta functions of \eqref{eq:F0onshell} and
\eqref{eq:F0boxonshell} allow us to solve the Bethe--Salpeter equation of \S\ref{sec:bse} by placing the rungs on shell.

\subsection{Rungs}
\label{sec:rungs}

\input{worked/ladder_on_loops_figure.tex} 

The diagrams that matter at large times are those leading in $\lambda$
and $N$ that can be stacked into a ladder; we call them rung functions.
Fig.~\ref{fig:ladderloops} shows the ladder they build on the surface the
correlator lives on.

Class (i) does not allow a single cubic rung: one exchanged gluon inserts
a single adjoint generator into each of the two colour cycles, and the
trace of a single adjoint generator vanishes (Appendix~\ref{app:colour}).
Its leading-order one-rung diagrams (without $c(t)$'s overall factor of $g^8=\lambda^4/N^4$ and the self-energy dressing of the rails) are of order $\lambda^2$, a result of either four cubic vertices, two quartic vertices, or one quartic and two cubic vertices. These diagrams are drawn in Fig.~\ref{fig:rungs} of Appendix~\ref{app:classirungs}. 

On the other hand, class (ii) allows a single cubic rung visible in Fig.~\ref{fig:rungbox}. As mentioned in \cite{stanford_many-body_2016}, bare cubic rungs cannot go on shell. Considering an HTL dressed cubic rung instead would have gotten it to order $\lambda^2$ for hard momenta $l \sim T$, because a single self energy insertion costs $\frac{1}{l^2}\Pi\frac{1}{l^2}$, and for $l \sim T$, $\Pi \sim m^2_{D} \sim \lambda T^2$ leading to $\lambda^2$. But, for soft momenta $l^2 \sim m^2_{D} \sim \Pi$, so $\frac{1}{l^2}\Pi \sim 1$. Meaning that in the kinematic regime we will be working in which is hard rails and soft rungs, the cubic ladder dressed by HTL will be of order $\lambda$, and can be placed on shell. We will explain what it means for the rung propagators in the following. To write down the one-rung functions, we will start by discussing the vertices and polarisation, continue with the loop integrals within the rungs and conclude with the one-rung functions themselves.

Consider first a cubic vertex on a rail: \eqref{eq:G3} contracted with
the polarisations and momenta of the rail it sits on.  On shell it
reduces to the eikonal factor $2\vec q$ of a fast charge, and we write
it as the vector $\vec V$.
On a rail carrying momentum $k$ with polarisation $\varepsilon$, the
vertex emits a gluon of momentum $\ell$, leaving the rail with $k-\ell$
and polarisation $\varepsilon'$.  Contract \eqref{eq:G3} with
$(p,q,r)=\big(k,-(k-\ell),-\ell\big)$ and with the two rail polarisations,
using transversality $\varepsilon\cdot\vec k=0$ and
$\varepsilon'\cdot(\vec k-\vec\ell\,)=0$, i.e.\
$\varepsilon'\cdot\vec k=\varepsilon'\cdot\vec\ell$.  The three terms of
\eqref{eq:G3} give
\begin{align}
g^{\mu\nu}(p-q)^\rho &\;\longrightarrow\;
(\varepsilon\cdot\varepsilon')\,(2\vec k-\vec\ell\,)\,,
\nonumber\\
g^{\nu\rho}(q-r)^\mu &\;\longrightarrow\;
\big[(-\vec k+2\vec\ell\,)\cdot\varepsilon\big]\,\varepsilon'
=2(\vec\ell\cdot\varepsilon)\,\varepsilon'\,,
\nonumber\\
g^{\rho\mu}(r-p)^\nu &\;\longrightarrow\;
\big[(-\vec\ell-\vec k)\cdot\varepsilon'\big]\,\varepsilon
=-2(\vec\ell\cdot\varepsilon')\,\varepsilon\,,
\nonumber
\end{align}
so that, stripping one power of $g$ and the colour factor $f^{abc}$,
\begin{equation}
\vec V(\vec k,\vec\ell\,)_{\varepsilon\varepsilon'}
=-\Big[(\varepsilon\cdot\varepsilon')(2\vec k-\vec\ell\,)
+2(\vec\ell\cdot\varepsilon)\,\varepsilon'
-2(\vec\ell\cdot\varepsilon')\,\varepsilon\Big]\,.
\label{eq:V}
\end{equation}
The absorbing vertex, where the rail takes the gluon in, is \eqref{eq:V}
with $2\vec k-\vec\ell\to2\vec k+\vec\ell$ and the same rotation terms.  The
first term is diagonal in polarisation and grows with the rail momentum; the
other two are smaller by $|\vec\ell\,|/2k$ and mix polarisations.  Since the
rails are hard ($k\sim T$) and the transfers soft
($\ell\sim\sqrt{\lambda}\,T$), we keep the diagonal term in the numerics
and treat the rotation terms as a $O(\sqrt{\lambda})$ correction.

Next, the quartic vertex. The quartic vertex has four legs: the rail 'in' ($a,\mu,\varepsilon$),
the rail 'out' ($b,\nu,\varepsilon'$), and the two exchanged gluons
($c,\rho$) and ($d,\sigma$).  Of the three colour structures in
\eqref{eq:G4}, the one with both $f$-indices on the rail legs,
$f^{abe}f^{cde}$, has vanishing colour on the loop's colour cycle, so only
the other two survive; they carry equal colour there and only the sum of
their index structures is needed.  Contracting with the rail polarisations,
\begin{equation}
\varepsilon_\mu\varepsilon'_\nu\Big[
\big(g^{\mu\nu}g^{\rho\sigma}-g^{\mu\sigma}g^{\nu\rho}\big)
+\big(g^{\mu\nu}g^{\rho\sigma}-g^{\mu\rho}g^{\nu\sigma}\big)\Big]
\;\Longrightarrow\;
W^{ij}(\varepsilon,\varepsilon')
=2(\varepsilon\cdot\varepsilon')\,\delta^{ij}
-\varepsilon^i\varepsilon'^j-\varepsilon^j\varepsilon'^i\,,
\label{eq:W}
\end{equation}
with $i,j$ handed to the two exchanged lines.  Note $W$ carries $g^2=\lambda/N$ and is
momentum independent, producing no
kinematic enhancement.

The remaining ingredient is the exchanged line.  Every exchanged gluon crosses one factor of $e^{-\beta H/2}$, so it carries
the Wightman function at imaginary-time separation $\beta/2$,
\begin{equation}
\mathcal G_{\mu\nu}(\ell)=
\frac{\rho_T(\ell)\,P^T_{\mu\nu}(\vec\ell\,)+\rho_L(\ell)\,P^L_{\mu\nu}(\ell)}
{2\sinh(\beta\ell^0/2)}\,,
\label{eq:rungline}
\end{equation}
in which both structures appear: the transfers are soft, and in $\xi = 0$
gauge the longitudinal piece is not removed.  It is the dressed
$\rho_{T,L}$ that matters, because the transfers are spacelike
(\S\ref{sec:bse}) and a bare gluon has no spectral weight there.  For
the two-gluon insertions of class (i) the dressing is not forced: two
internal lines can sit on shell with a spacelike total, one future-null
and one past-null; we keep the dressed lines throughout for uniformity.
The dressing costs no power of the coupling.  The transfer is soft,
$\ell\sim m_D$, where the hard-thermal-loop self-energy is
$\Pi\sim m_D^{2}\sim\ell^{2}$, so each self-energy insertion is the same
size as the one before it and the series must be summed rather than
truncated \cite{braaten_soft_1990}; the resummed line is the
leading-order propagator for a soft gluon, not a correction to one.  This
also settles what happens to the exchanged gluon's own width.  A rail sits
at its pole, where a width is what smears it, and that is the width the
loop's size cuts.  An exchanged gluon never reaches its pole: its frequency
is fixed by \eqref{eq:quasielastic} and stays spacelike.  What it carries
instead is the imaginary part of the dressing itself, which is not a decay width in this kinematic regime
and is unaware of the loop size because it has no endpoints on it.

\begin{figure}[htbp]
\centering
\rungbox{\dKbox}
\caption{The class-(ii) rung: a single dressed gluon exchanged between the
two rails $A$ and $B$ of class (ii).  Here the adjoint trace does not
vanish, so one gluon is allowed and the rung enters at $O(\lambda)$.}
\label{fig:rungbox}
\end{figure}

Class (ii) is the simpler of the two.  Colour permits a single exchanged
gluon, so the rung is one line \eqref{eq:rungline} between the two rails,
with one cubic vertex at each end,
\begin{equation}
R^{(ii)}\big(\vec q\,;\ell\big)
=-\,\lambda\;
\vec V(\vec q,\vec\ell\,)\cdot\mathcal G(\ell)\cdot
\vec V(-\vec q,-\vec\ell\,)\,,
\label{eq:Rbox}
\end{equation}
\bigskip

Class (i) is more complicated.  Its two-gluon insertions fall into six families, with $64$
placements. The enumeration, the diagrams, the
colour sign each placement carries and the conventions common to all of
them are collected in Appendix~\ref{app:classirungs}.  We display one
family per vertex content here.
\filbreak
\bigskip

Four cubic vertices and no internal loop.
\begin{center}
\rungwide{\raildmom{k_1'}{k_2'}{k_3'}{k_4'}\dKone}
\end{center}
\nopagebreak
\begin{align}
R^{(1)}\big(\{k\};\ell_1,\ell_2\big)
=\Phi_4\,g^4\;
&\vec V(\vec q_1,\vec\ell_1)_{\varepsilon_1\varepsilon_1'}\!\cdot
\mathcal G(\ell_1)\cdot
\vec V(\vec q_3,-\vec\ell_1)_{\varepsilon_3\varepsilon_3'}
\nonumber\\[-2pt]
&\times\;
\vec V(\vec q_2,\vec\ell_2)_{\varepsilon_2\varepsilon_2'}\!\cdot
\mathcal G(\ell_2)\cdot
\vec V(\vec q_4,-\vec\ell_2)_{\varepsilon_4\varepsilon_4'}\,,
\label{eq:R1}
\end{align}
a pure function of the rail momenta and the two transfers: gluon~1
shifts rails 1 and 3 by $\mp\ell_1$, gluon~2 shifts rails 2 and 4 by
$\mp\ell_2$, and the function carries no integral.

\bigskip

\bigskip

\bigskip

\bigskip

\noindent Two four-gluon vertices, one internal loop, and a symmetry
factor $\tfrac12$.  

\begin{center}
\rungwide{\raildmom{k_1'}{k_2'}{k_3'}{k_4'}\dKfour}
\end{center}
\nopagebreak
\begin{multline}
R^{(4)}\big(\{k\};L\big)=\tfrac12\,\Phi_2\,g^4\!
\int\!\frac{\dd^4\ell}{(2\pi)^4}\;
W^{ij}(\varepsilon_1,\varepsilon_1')\,
\mathcal G_{ii'}(\ell)\,\mathcal G_{jj'}(L{-}\ell)\,
W^{i'j'}(\varepsilon_3,\varepsilon_3')\,,
\label{eq:R4}
\end{multline}

\bigskip

\eqref{eq:R1} is the all-cubic family and \eqref{eq:R4} the all-quartic
one.  The other four are written out in Appendix~\ref{app:classirungs}; together with the mirror placements they
exhaust the $64$ labelled placements, and with \eqref{eq:V},
\eqref{eq:W}, \eqref{eq:rungline} and \eqref{eq:coloursign} they
determine the kernel completely.

\FloatBarrier
\subsection{The Bethe--Salpeter equation}
\label{sec:bse}

\begin{figure}[htbp]
\centering
\begin{tikzpicture}[xscale=0.68,yscale=0.46]
\node at (-5.4,1.5) {(i)};
\begin{scope}[shift={(-4.0,0)}]
  \foreach \y in {3,2.2,0.8,0} \draw[thick] (0,\y)--(2.8,\y);
  \draw[thick,fill=white] (0.6,-0.5) rectangle (2.2,3.5);
  \node at (1.4,1.5) {$F$};
\end{scope}
\node at (0.2,1.5) {$=$};
\begin{scope}[shift={(1.1,0)}]
  \foreach \y in {3,2.2,0.8,0} \draw[thick] (0,\y)--(2.6,\y);
  \node at (1.3,-1.1) {\footnotesize $F_0$};
\end{scope}
\node at (4.5,1.5) {$+$};
\begin{scope}[shift={(5.5,0)}]
  \foreach \y in {3,2.2,0.8,0} \draw[thick] (0,\y)--(2.3,\y);
  \node at (1.15,-1.1) {\footnotesize $F_0$};
  \foreach \y in {3,2.2,0.8,0} \draw[thick] (2.3,\y)--(5.1,\y);
  \draw[thick,dashed] (2.3,-0.5) rectangle (4.5,3.5);
  \draw[gl] (2.9,3)--(2.9,0.8);
  \draw[gl] (3.9,2.2)--(3.9,0);
  \node at (3.4,-1.1) {\footnotesize $R$};
  \draw[thick,fill=white] (5.1,-0.5) rectangle (6.7,3.5);
  \node at (5.9,1.5) {$F$};
\end{scope}
\end{tikzpicture}

\vspace{5mm}

\begin{tikzpicture}[xscale=0.68,yscale=0.46]
\node at (-5.4,1.5) {(ii)};
\begin{scope}[shift={(-4.0,0)}]
  \foreach \y in {3,0} \draw[thick] (0,\y)--(2.8,\y);
  \draw[thick,red!70!black] (0,3) .. controls (-0.7,2.1) and (-0.7,0.9) .. (0,0);
  \draw[thick,red!70!black] (2.8,3) .. controls (3.5,2.1) and (3.5,0.9) .. (2.8,0);
  \draw[thick,fill=white] (0.6,-0.5) rectangle (2.2,3.5);
  \node at (1.4,1.5) {$F$};
\end{scope}
\node at (0.2,1.5) {$=$};
\begin{scope}[shift={(1.1,0)}]
  \foreach \y in {3,0} \draw[thick] (0,\y)--(2.6,\y);
  \draw[thick,red!70!black] (0,3) .. controls (-0.7,2.1) and (-0.7,0.9) .. (0,0);
  \draw[thick,red!70!black] (2.6,3) .. controls (3.3,2.1) and (3.3,0.9) .. (2.6,0);
  \node at (1.3,-1.1) {\footnotesize $F_0$};
\end{scope}
\node at (4.9,1.5) {$+$};
\begin{scope}[shift={(5.9,0)}]
  \foreach \y in {3,0} \draw[thick] (0,\y)--(2.3,\y);
  \draw[thick,red!70!black] (0,3) .. controls (-0.7,2.1) and (-0.7,0.9) .. (0,0);
  \node at (1.15,-1.1) {\footnotesize $F_0$};
  \foreach \y in {3,0} \draw[thick] (2.3,\y)--(4.6,\y);
  \draw[thick,dashed] (2.3,-0.5) rectangle (4.0,3.5);
  \draw[gl] (3.15,3)--(3.15,0);
  \node at (3.15,-1.1) {\footnotesize $R$};
  \draw[thick,fill=white] (4.6,-0.5) rectangle (6.2,3.5);
  \node at (5.4,1.5) {$F$};
  \draw[thick,red!70!black] (6.2,3) .. controls (6.9,2.1) and (6.9,0.9) .. (6.2,0);
\end{scope}
\end{tikzpicture}
\caption{The Bethe--Salpeter equation \eqref{eq:BSE} in both channels: the
free block, plus the free block followed by one insertion $R$ and then the
full four-point function.  Top, class (i): four rails, and $R$ is the sum
over all placements of Fig.~\ref{fig:rungs}, two exchanged gluons drawn
schematically inside the dashed box.  Bottom, class (ii): two rails, and
$R$ is the single dressed gluon \eqref{eq:Rbox}; the two equal-time caps
are drawn in red at the ends of the interval.  They appear identically on both sides of the equation through the free
block and divide out, which is why the class-(ii) kernel acts on one
momentum instead of two.}
\label{fig:bse}
\end{figure}

Let $R$ denote the one-rung kernel, indexed by family $j$ and placement $i$
within it,
\begin{equation}
R=\sum_j\sum_i R_{ji}\,,
\label{eq:Rsum}
\end{equation}
each $R_{ji}$ built from the vertices \eqref{eq:G3}--\eqref{eq:G4}, the rung lines
\eqref{eq:rungline}, and (where the placement has one) an internal loop
integral.  The resummation is
\begin{equation}
F=F_0+F_0\,R\,F\,,
\label{eq:BSE}
\end{equation}
drawn in Fig.~\ref{fig:bse} for both classes:
the action of $R$ integrates each placement's transfers, one
$\dd^4\ell/(2\pi)^4$ each, against $F$ at the shifted rail momenta its
diagram displays; the rails a placement does not touch keep theirs.

Write \eqref{eq:BSE} out for class (i) with $F_0^{(i)}$ in place,
\begin{equation}
F(\{k\})=\Delta^R_T(k_1)\,\widetilde D_T(k_2)\,\Delta^R_T(k_3)\,
\widetilde D_T(k_4)\,
\Big[1+\big[R\,F\big](\{k\})\Big]\,,
\label{eq:BSEi}
\end{equation}
and for class (ii) the same with $F_0^{(ii)}$, two rails instead of four and
the equal-time contractions as overall factors. We drop the homogeneous term because it decays and in large times does not contribute.  From here
on we also write $f(\{k\};\omega)$ for $F$ with the class (ii) equal-time
contractions set aside: they close the two ends of the ladder, so they
enter the solution once, as an overall factor, and do not affect the exponent.

Now substitute the on-shell forms \eqref{eq:F0onshell} and
\eqref{eq:F0boxonshell} for every free block in the series; as in
\S\ref{sec:free}, this keeps the part that governs large times.    Multiplying through by the pole denominator then turns
\eqref{eq:BSEi} into
\begin{multline}
-i\omega\,f(\{k\};\omega)=-\sum_j\gamma(q_j)\,f(\{k\};\omega)
-2\pi\!\!\sum_{s_1,s_3=\pm}\!\!s_1s_3\,
\frac{Z_T(q_1)}{2\omega_T(q_1)}\,\frac{Z_T(q_3)}{2\omega_T(q_3)}\,
\delta\big(k_1^0-s_1\omega_T(q_1)\big)\\
\times\prod_{j=2,4}2\pi Z_T(q_j)\Big[n_B(\omega_T(q_j))+\tfrac12\Big]
\delta\big((k_j^0)^2-\omega_T^2(q_j)\big)
\big[R\,f\big](\{k\};\omega)
\label{eq:onshellBSEi}
\end{multline}
for class (i), and
\begin{multline}
-i\omega\,f(k_A;\omega)=-2\gamma(q)\,f(k_A;\omega)
+2\pi\Big[\frac{Z_T(q)}{2\omega_T(q)}\Big]^{2}
\Big[\delta\big(k_A^0-\omega_T(q)\big)+\delta\big(k_A^0+\omega_T(q)\big)\Big]\\
\times\;\big[R\,f\big](k_A;\omega)
\label{eq:onshellBSEii}
\end{multline}
for class (ii).  The deltas of \eqref{eq:F0onshell} and
\eqref{eq:F0boxonshell} fix every rail frequency to its dispersion,
$k_j^0=s_j\,\omega_T(q_j)$.  Energy conservation at the two vertices of
an exchange then fixes its frequency to the difference of the two rail
frequencies,
\begin{equation}
\ell^0=\omega_T(q)-\omega_T\big(|\vec q-\vec\ell\,|\big)\,.
\label{eq:quasielastic}
\end{equation}
The branch with the sum of the two frequencies sits near $2\omega_T$ and
has no soft spectral weight there.  The difference branch obeys
$|\ell^0|<|\vec\ell\,|$, because the group velocity is below one, so
every exchange is spacelike.  A bare gluon has no spectral weight at
spacelike momenta; the dressed one does, because Landau damping,
scattering off the hard particles of the bath, gives $\rho_{T,L}$
spacelike support.  That support is called the Landau cut, and it is all
the weight the rungs have.  

This leaves an equation for a function of
the spatial momenta alone: $f(\omega;q_1,q_3,x)$ with
$x=\hat q_1\cdot\hat q_3$ for class (i), $f(\omega;q)$ for class (ii);
\eqref{eq:l0delta} carries one case out in full.  Each rung function of \S\ref{sec:rungs} then becomes an operator
$\mathcal K_r$ on $f$. 
Appendix~\ref{app:kernels} writes the class (ii) reduction out in full. In
the sector \eqref{eq:pinch}, $\sum_j\gamma(q_j)=2\gamma(q_1)+2\gamma(q_3)$.  For class (i),
\begin{equation}
\boxed{
\;-i\omega\,f(\omega;q_1,q_3,x)=
-\big[2\gamma(q_1)+2\gamma(q_3)\big]f(\omega;q_1,q_3,x)
+\sum_{r=1}^{64}\big[\mathcal K_r f\big](\omega;q_1,q_3,x)\;}
\label{eq:finali}
\end{equation}
the sum running over every labeled placement, that is the six rows of
Fig.~\ref{fig:rungs} with their multiplicities $4,16,16,4,8,16$, which is
where the $64$ comes from.  Placements are summed rather than selected, since the signs of
\eqref{eq:coloursign} are not uniform within a family.
For class (ii), where the single exchanged momentum leaves a one-dimensional state,
\begin{equation}
\boxed{\;-i\omega\,f(\omega;q)=-2\gamma(q)\,f(\omega;q)
-\,\lambda\!\int\!\frac{d^3\ell}{(2\pi)^3}\Big[\frac{Z_T(q')}{2\omega_T(q')}\Big]^2
\big[\vec n\cdot\mathcal G(\ell^0,\vec\ell\,)\cdot\vec n\,'\big]f(\omega;q')\;}
\label{eq:finalbox}
\end{equation}
with $\ell^0$ fixed by \eqref{eq:quasielastic}.  We Laplace transform back to real time, and solve the eigenvalue equation numerically. Each 
eigenvalue $m$ contributes $e^{mt}$ after the Laplace transform, so
\begin{equation}
\lambda_L=\max_m\ \mathrm{Re}\,m\,.
\label{eq:lamL}
\end{equation}
The rates in the equation depend on the momenta; the momentum
dependence of the solution sits in the eigenfunctions, and at large
times the eigenvalue with the largest real part dominates, which is
\eqref{eq:lamL}.  The full OTOC $c(t)$ is then the contour integration
of \eqref{eq:master} acting on the eigenfunction of $\lambda_L$.
 The ladder summed above is built from one kind of insertion: a rung crossing between two thermal folds along the imaginary time contour. There is a
second kind, drawn on the right of Fig.~\ref{fig:samerail}, in which the
exchanged gluon connects the two lines belonging to the same commutator,
that is, to the same fold of the time contour. It is a bit peculiar, because its thermal propagation is over a width of $i\epsilon$ in imaginary time, not
by the finite $\beta/2$ that separates the two commutators and fixes the
thermal weight every rung above carries. We do not know if it contributes.

\input{worked/samerail_figure.tex}

\bigskip

\subsection{The Lyapunov exponent}
\label{sec:lyap}

\subsubsection{Assumptions}

Equation \eqref{eq:finalbox} is discretised on a geometric grid in the
rail momentum, the transfer integral evaluated on the HTL spectral
functions, and the resulting matrix is diagonalised.  We give the
numerical detail in Appendix~\ref{app:numerics}.

We solve class (ii) only.  Class (i) ladders are complicated to compute.
They contain seven rung families and sixty-four labelled placements, and
we found no obvious cancellation among them.  A rung also moves momentum
between all four rails, so after one rung the two rails of a pair are no
longer back to back and the state needs another variable to say how far
off they are.  Although class (i) is the leading class in colour, its
rungs carry a power $\lambda^2$ against the dressed class (ii) rung's
$\lambda$.  We quote no exponent for it and cannot say how its growth
rate compares with class (ii)'s; we leave it for future investigation,
and the rung functions we identified are given in full in
Appendix~\ref{app:classirungs}.

We make the following assumptions.

(a) Zero total momentum.  In general $\lambda_L$ depends on the
momentum injected into the ladder.  As in both \cite{stanford_many-body_2016} and \cite{steinberg_thermalization_2019}: the equations above are with
$k=0$.  This is a simplifying assumption, since we have not computed $\lambda_L(k)$.

(b) The width for a hard transverse gluon is given by
\cite{pisarski_moving_1993,flechsig_rebhan_schulz_1995}
\begin{equation}
\gamma(q)\;=\;\frac{\lambda}{4\pi}\,v_t(q)\,
\Big[\ln\frac{m_D}{m_{\rm mag}}+0.548\Big]\,,
\qquad v_t(q)=\frac{d\omega_T}{dq}\,,
\label{eq:gammafull}
\end{equation}
with $v_t$ the group velocity of the transverse branch, given in closed
form in \cite{flechsig_rebhan_schulz_1995}; our rails are hard,
$q\in[0.5,4]\,T$, where the group velocity is $v_t=0.96$--$1.00$, so
\eqref{eq:gammafull} is the standard hard-gluon rate.  The log counts the unscreened magnetic exchanges between the
magnetic scale and the Debye mass, and the constant $0.548$ is the
contribution of the exchanges at the screening scale itself
(eq.~(4.8) of \cite{pisarski_moving_1993}). Both masses scale with
$\lambda$, so at large $N$ the log reads $\tfrac12\ln(1/\lambda)$ up
to a pure number.  The IR end needs no cutoff by hand, the magnetic scale supplies it;
but that scale is not perturbative, it is where the magnetostatic sector
becomes strongly coupled \cite{linde_infrared_1980}, and
\eqref{eq:gammafull} inherits exactly that sensitivity through the lower
limit of its log.

We explore the possibility of replacing that limit by the loop's own scale, as a physical IR cutoff. To truly achieve an accurate estimate for $\lambda_{L}$, further investigation into the width is required. Regardless, the separation of any two points on the loop is bounded by
the diameter, $d\le2R$,
so exchanges with $\ell\lesssim1/2R$ are invisible to the loop correlators.  We therefore replace the IR end of the log by
$1/2R$,
\begin{equation}
\gamma(q;R)\;=\;\frac{\lambda}{4\pi}\,v_t(q)\,
\Big[\ln\big(2R\,m_D\big)+0.548\Big]\,,
\label{eq:gammacut}
\end{equation}
with the log set to zero when $2R\,m_D<1$, and capped at $\tfrac12\ln(1/\lambda)$,
where the magnetic scale takes over from the loop.  The constant term is kept for all the numbers we report, so that loss
treats the screening-scale exchanges the same way the kernel's gain
does.  A calculation of the rails' damping in the loop's presence would remove
it; for static sources an analogous calculation exists
\cite{laine_realtime_2007}.

\input{worked/cutoff_geometry_figure.tex}
The cutoff applies to the rails and not to the rungs, and
Fig.~\ref{fig:cutoffgeom} shows why.  A rail's damping
comes from its exchanges with the plasma, and the two rails of one
commutator carry opposite colour charge: an exchange longer than their
separation sees no net charge, and that is what the cut removes.  A rung
runs between the two commutators; it has no partner to cancel against,
so its transfer integrals are kept in full, down to the magnetic scale,
for which we take $\sqrt{\lambda}\,m_D$.  The cutoff therefore acts on
one term of the equation and not on the other, and this is why the
answer depends on $R$.

(c) The on-shell reduction.  Every free block in the series is replaced by
its on-shell form, \eqref{eq:F0onshell} and \eqref{eq:F0boxonshell}, and
the exchanged frequencies are fixed by \eqref{eq:quasielastic}.  This keeps
the part of the correlator that governs large times; what it drops is down
by $\gamma/\omega_T$, the same approximation as \eqref{eq:R13}.

(d) A rung gluon never reaches its pole: its frequency is fixed by
\eqref{eq:quasielastic} and is spacelike, on the Landau cut, so its entire
weight is the imaginary part of the self-energy dressing it, the resummed
propagator whose soft-frequency form is given in appendix~B of
\cite{laine_realtime_2007}.

\subsubsection{Results and the transition}

\begin{figure}[htbp]
\centering
\includegraphics[width=\textwidth]{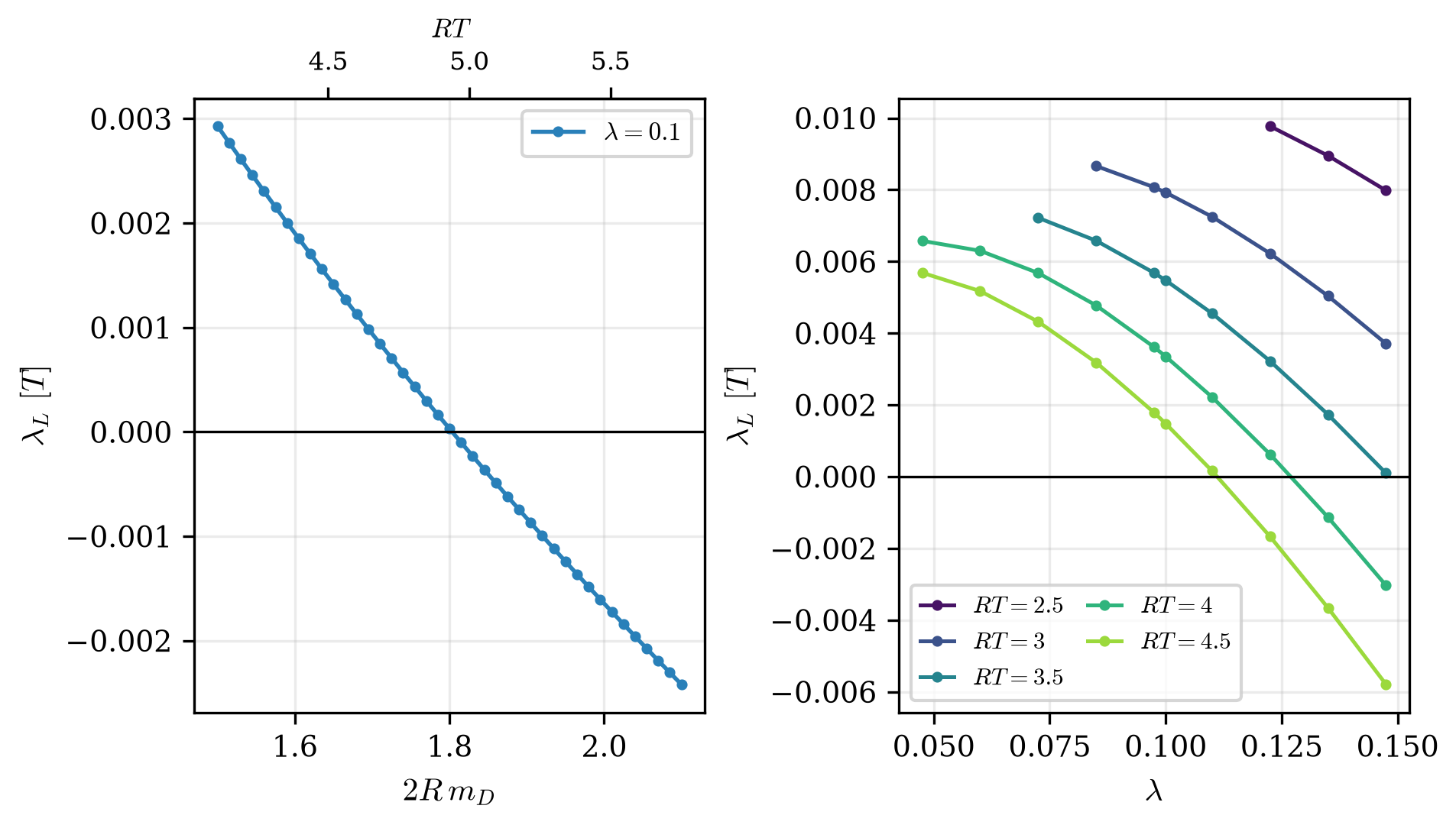}
\caption{Left: the Lyapunov exponent of the loop commutator squared at
$\lambda=0.1$, against the loop diameter in units of the Debye length,
on a window around the zero crossing at $2R^*m_D\simeq1.8$; the whole
window lies inside the region $2R\,m_D\ge1$ where the width
\eqref{eq:gammacut} is valid.  The upper axis converts to thermal units.
Right: the same exponent against the coupling at
fixed loop radius $RT$, each curve drawn from the coupling at which its
loop reaches the Debye length, $\lambda=3/(2RT)^{2}$, upward.  Both
panels: class (ii), width \eqref{eq:gammacut}, zero total momentum.}
\label{fig:master}
\end{figure}

\begin{figure}[htbp]
\centering
\includegraphics[width=0.72\textwidth]{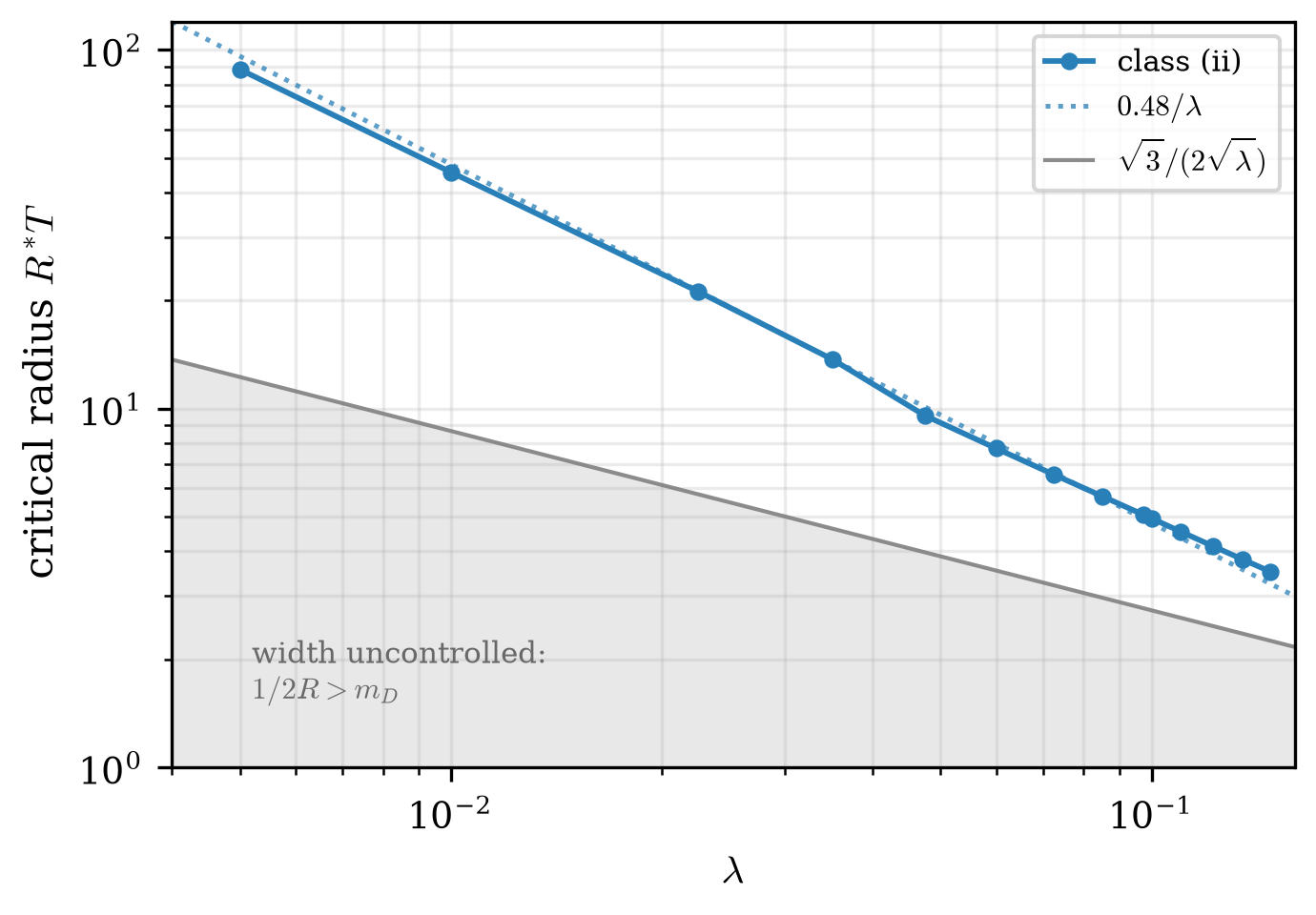}
\caption{The critical loop radius in thermal units.  Class (ii) follows
\eqref{eq:rcritfit}, the dotted guide being $0.48/\lambda$; the grey line
is the Debye length, $RT=\sqrt3/(2\sqrt\lambda)$, and the shaded region
below it is $1/2R>m_D$, where the width expression is not valid and no
exponent is reported.}
\label{fig:rcrit}
\end{figure}

First, the calculation with the width as it stands and no reference to the
loop anywhere: \eqref{eq:gammafull} on the diagonal, the kernels as built. Without any IR cutoff, the width
carries the magnetic scale in the lower limit of its log, and the
kernels carry it at the soft end of their transfer integrals: numerically, the kernel eigenvalue moves with the log of that soft end as it is lowered,
and nothing perturbative stops it \cite{linde_infrared_1980}.  So we cannot compute the exponent of the loop commutator without calculating the decay width with all its IR sensitivity, which we have not done here. Whether the plasma's own magnetic sector organises growth or decay is an open question. To explore how the exponent would behave with an IR cutoff, we assume \eqref{eq:gammacut}: the loop size itself supplies the
IR end.  Its domain must be stated first: for $2R\,m_D<1$ the cut $1/2R$
lies above the Debye mass, while the damping \eqref{eq:gammafull}
describes exchange below $m_D$, so the width expression stops being valid there and we report no exponent; every figure of this subsection is drawn on $2R\,m_D\ge1$. Figure~\ref{fig:master} shows the outcome at $\lambda=0.1$, and it is a transition in loop size: the
channel grows below a critical diameter and decays above it,
\begin{equation}
2R^*_{(ii)}\,m_D\simeq1.8
\qquad\text{at}\qquad \lambda=0.1\,,
\label{eq:Rstar}
\end{equation}
and Figure~\ref{fig:rcrit} follows that critical size across the coupling
range $0.005\le\lambda\le0.15$.  The crossing lies well inside the region
of validity, with its critical radius fitted to
\begin{equation}
2R^*_{(ii)}\,m_D\;=\;0.56\,\lambda^{-1/2}\,,
\qquad\text{which is}\qquad
R^*_{(ii)}T\;\approx\;0.48/\lambda\,,
\label{eq:rcritfit}
\end{equation}
numerically consistent with the magnetic length over the range fitted (a
free fit returns exponent $-0.46$; the fixed $-\tfrac12$ form holds to
$\pm10\%$ over $0.005\le\lambda\le0.17$).

What we consider robust is the pattern: a loop-size threshold for
scrambling whose location follows an IR length of the plasma,
numerically the magnetic length.

The exponent crosses zero continuously at $R^*$, so the scrambling time
diverges there. This is the familiar
critical slowing down of a continuous dynamical transition: loops smaller
than $R^*$ scramble in a finite time, loops larger than $R^*$ never do.  The saturation value itself is fixed by
large-$N$ factorisation of the four-point function.  What the exploration
establishes is a loop-size threshold for scrambling at weak coupling,
tied to an IR length of the plasma screening, intuitively dictating
whether perturbations die down or diverge.

To compute the full four-point function $c(t)$, the eight contour integrals of \eqref{eq:master} act on the eigenfunction of $\lambda_L$. For a planar circular loop the tangential current is
\begin{equation}
J^\mu(\vec q\,)=\oint ds\,\dot x^\mu(s)e^{i\vec q\cdot x(s)}
=-2\pi iR\,J_1(q_\perp R)\,\hat\varphi^\mu\,,
\label{eq:Jcurrent}
\end{equation}
which is orthogonal to $\vec q$, so each rail's transverse projector acts on
it as the identity and each rail contributes $(2\pi R)^2J_1(q_\perp R)^2$.
The correlator for class (i) is
\begin{equation}
c(t;R)=\mathcal N_{(i)}\int\prod_{j=1}^{4}
\Big[\frac{\dd^3q_j}{(2\pi)^3}\,(2\pi R)^2J_1(q_{j\perp}R)^2\Big]\;
\mathcal P(q_1,\dots,q_4)\;f\big(t;q_1,q_3,x\big)\,,
\label{eq:assembled}
\end{equation}
where the prefactor is the one carried down from \eqref{eq:master}
together with the free colour factor of the channel.  Counting it out,
\begin{equation}
\mathcal N_{(i)}=\frac{\lambda^4}{256\,N^8}\cdot\frac{(N^2-1)^2}{16}
\;\xrightarrow[N\to\infty]{}\;\frac{\lambda^4}{4096\,N^4}\,,
\label{eq:Ncount}
\end{equation}
and for class (ii) the same with $(N^2-1)/16$, giving $\lambda^4/4096N^6$.
The loop out-of-time-order correlator is therefore $O(N^{-4})$ and not
$O(N^{-2})$, which is the counting quoted for generic single-trace
operators.  The reason is that a normalised loop is not a generic operator:
expanding it to its leading non-trivial order already costs $\lambda/N^2$,
so each of the four loops contributes one power of $\lambda/N^2$, and the
colour cycles return only $N^4$ of the $N^8$. Here $q_{j\perp}$ is the part of $\vec q_j$ in the loop plane,
$x=\hat q_1\cdot\hat q_3$, $\mathcal P$ is the weight each rail carries
once it is on shell, and $\mathcal N$ is an $R$- and $t$-independent
constant. $\mathcal P$ is
one factor per rail, the residue of \eqref{eq:pfrac} for the two retarded
rails and the same residue times the occupancy of
\eqref{eq:symmetric} for the two thermal ones,
\begin{equation}
\mathcal P(q_1,\dots,q_4)=
\prod_{r=1,3}\frac{Z_T(q_r)}{2\omega_T(q_r)}\;
\prod_{r=2,4}\frac{Z_T(q_r)}{2\omega_T(q_r)}
\Big[n_B\big(\omega_T(q_r)\big)+\tfrac12\Big]\,,
\label{eq:Pdef}
\end{equation}
The four momenta are independent: on a fixed contour nothing ties
$\vec q_2$ to $-\vec q_1$, and each rail carries its own $J_1^2$ weight.
Those weights support each pair's total momentum, $\vec q_1+\vec q_2$ and
$\vec q_3+\vec q_4$, only up to $\sim1/R$, and on that support $f$ is
evaluated in the zero-momentum block of assumption (a) of
\S\ref{sec:lyap}, where \eqref{eq:backtoback} holds and the arguments of
$f$ are $q_1$, $q_3$ and $x$.

The class-(ii) case has one rail pair and two caps.  Each cap is its
own contraction, with its own momentum and its own $J_1^2$ weight; it
carries no time dependence, so it enters as an $R$-dependent constant,
one per cap,
\begin{equation}
C_{\rm cap}(R)=\int\!\frac{\dd^3q_c}{(2\pi)^3}\,(2\pi R)^2
J_1(q_{c\perp}R)^2\,B(q_c)\,,
\label{eq:capdef}
\end{equation}
with $B$ the cap function of \eqref{eq:F0box}, and
\begin{equation}
c_{(ii)}(t;R)=\mathcal N_{(ii)}\,C_{\rm cap}(R)^2
\int\prod_{r=A,B}\Big[\frac{\dd^3q_r}{(2\pi)^3}\,(2\pi R)^2
J_1(q_{r\perp}R)^2\,\frac{Z_T(q_r)}{2\omega_T(q_r)}\Big]\,
f\big(t;q_A\big)\,,
\label{eq:assembledbox}
\end{equation}
with $f$ on the zero-momentum block, as for class (i).  The caps multiply
both sides of \eqref{eq:BSE} identically and divide out of the iteration,
which is why they enter only here and never the exponent.
We do not settle which channel carries the
larger amplitude, as it depends not only on the colour counting, but also on the momentum integrals \eqref{eq:assembled}
and \eqref{eq:assembledbox} as well.  The
two carry different $R$-dependent weights, and each is evaluated on an
eigenfunction whose normalisation the eigenvalue problem does not fix, so
their ratio is not a pure number that could be weighed against $N^2-1$.
In the correlator the loop's finite size is
already present, through the current \eqref{eq:Jcurrent}:
$J_1(q_\perp R)$ is the Fourier transform of the contour itself, so the
suppression of wavelengths longer than the loop is a form factor rather
than a cutoff. The growth region is also bounded from below, by two effects the
fixed-coupling calculation does not contain. Once $1/R$ exceeds the thermal scale, the coupling itself
runs down at the loop's scale, so ever smaller loops are ever more weakly
coupled probes.  Both effects switch on for $RT\lesssim1$.  The growth
region is therefore a window in loop size, from roughly the thermal
wavelength up to the critical sizes of \eqref{eq:Rstar}, with the maximum
exponent set by the coupling at the thermal scale.  It would be interesting to calculate the running of the exponent with its beta function.

\section{Strong coupling}
\label{sec:strong}
\label{sec:cobordism}

At strong coupling the computation of Sec.~\ref{sec:weak} has no bearing and gluons are not considered to be the fundamental excitations. We consider using the
framework of Sec.~\ref{sec:dynamics} to describe the loop evolution at the strongly coupled regime. The loop operator evolves
by displacement of its contour, and the events that change the operator's
character are the self-crossings at which the loop cuts. Those events are thought to
happen at a fixed physical scale, $\ell_s\sim\Lambda_{\rm QCD}^{-1}$, not
at a scale set by the coupling.  This section is a proposal: what the two-point and four-point functions of loops might be as ``worldsheet'' objects,
and what needs to be calculated to compute a Lyapunov exponent for example. This worldsheet description assumes loops large and
smooth on the scale of the string length. In the strongly coupled regime a natural object interpolating between two loops
is a surface bounded by them.  A compact orientable surface whose boundary is the disjoint union of the two contours is by definition a cobordism between them, and for a connected surface with a fixed number of boundaries the diffeomorphism classes are labelled by one integer, the genus. Two proposals follow. 

The first is that the connected two-point function of loops is organised as
a sum over such cobordisms graded by genus,
\begin{equation}
\big\langle W^t_\gamma W^{0\dagger}_\gamma\big\rangle_c
=\sum_{g\ge0} N^{-2(g+1)}\,d_g\,,
\label{eq:cobordism}
\end{equation}

\input{worked/cobordisms_figure.tex}
the genus-zero term being the cylinder, the loop propagator itself, and
each higher term adding a handle, as in Fig.~\ref{fig:cobordisms}.

The topology is fixed by elementary counting. A surface of genus $g$ with
$n$ boundaries decomposes into $p=2g-2+n$ pairs of pants; at $n=2$ that
is $p=2g$, so $g$ splittings and $g$ joinings, and the Euler
characteristic is $\chi=-2g$.  A genus-$g$ cobordism with two boundaries
weighs $N^{-2(g+1)}$, which is what appears in \eqref{eq:cobordism}; the
string coupling this expansion identifies is fixed below, once a handle's
cut content is counted, to $g_s=\varepsilon/N$.

Each cut event carries a weight, call it $\varepsilon$; at weak coupling
\S\ref{sec:growth} prices a cut at $t^{2}\lambda$, and at strong coupling
we do not know it. A genus-$g$ surface contains $2g$ of them, so
\begin{equation}
d_g\;\sim\;\varepsilon^{2g}\,d_0\,,
\label{eq:dgeps}
\end{equation}
and the dressing of the cylinder runs in even powers of $\varepsilon$,
because a handle is one split followed by one join and never costs an odd
number of cuts.  The two expansions, in $1/N^2$ and in $\varepsilon$, are
therefore not independent: they are the same pants graded two ways, with
$n=p+2-2g$.

The genus of the cobordism is not the genus of the 't~Hooft index loop, the
first counts splittings of the physical loop, the second counts
the $1/N^{2}$ suppression of the colour contractions. A cobordism of fixed genus therefore still carries its own $1/N^2$ series of index dressings, $d_g=d_g^{(0)}+d_g^{(1)}/N^2+\dots$, and the two gradings must be kept
apart.  Two-dimensional Yang--Mills is a sanity check for such a framework
\cite{witten_2dym_1991,gross_taylor_1993}: there the source of
\eqref{eq:source} vanishes identically, as \S\ref{sec:growth} shows, so
there are no cuts at all, the cobordism genus never leaves zero while the index expansion is nontrivial, and \eqref{eq:cobordism} collapses to its first term.

The four point function carries four single-trace insertions, so within the same organisation it is a $2\to2$ 
closed-string scattering (see Figure~\ref{fig:genusexp} ) amplitude: at genus zero the four-holed sphere,
$\chi=-2$, built from two pairs of pants glued on a common circle.  Two
loops enter at time $0$, two leave at time $t$, and the ladder of
Sec.~\ref{sec:weak} is the exchanged channel.

\input{worked/genus_expansion_figure.tex}
A handle costs one
factor $N^{-2}$ from the 't~Hooft counting and one factor $\varepsilon^{2}$
from the two consecutive Makeenko--Migdal cuts it contains, so
$g_s^{2}=N^{-2}\varepsilon^{2}$: the string coupling of this expansion is
$g_s=\varepsilon/N$, as identified with \eqref{eq:cobordism}.

Turning this into a number would need the four-string amplitude in a
confining background.  Effective string theory
\cite{aharony_komargodski_2013} does not supply it, being a
long-wavelength expansion around the static flux tube, and neither does
the loop equation as it stands: the Makeenko--Migdal contact term at a
self-intersection needs a regularisation, and lattice, dimensional and
point-splitting schemes differ in what they assign to it
\cite{makeenko_methods_2002,makeenko_notes_2025}.  Fixing that scheme is
what would turn the identification of a cut with a pair of pants into a
calculation, and the same amplitude would give the time at which the loop
splits, from Fig.~\ref{fig:pants}(b).

\FloatBarrier
\section{Discussion and outlook}
\label{sec:discussion}

Based on our analysis of Sec.~\ref{sec:dynamics}, operators in pure-gauge
Yang--Mills grow only when the gauge group is non-abelian and the number
of spacetime dimensions exceeds two, and we conjecture that pure gauge theories
are chaotic only when non-abelian at $d>2$.  The question is then
quantitative: how fast, and for which operators.

We attempted to answer that question at weak coupling for the simplest non-local gauge-invariant operator, a circular Wilson loop of radius $R$.  Expanding the loop commutator to leading order turns the out-of-time-order correlator into a gluon four-point function; of its $36$ surviving
contractions, two topologies admit an iterable ladder.  We solve the
class (ii) equation numerically and write down, but do not solve, the
class (i) one. The colour algebra distinguishes them sharply: a single exchanged gluon vanishes identically between the two commutators of the within-commutator topology, so its ladder starts at $O(\lambda^2)$, while class (ii) permits the single-gluon rung and competes with the rails' damping at the same order in $\lambda$.

We learned that the Lyapunov exponent is IR-sensitive, through the
magnetic scale in the HTL self-energy imaginary part, and through the soft end of the ladder's transfer integrals. We also learned that imposing the loop size as an IR cutoff acting as
the magnetic scale leads to a transition, at the critical diameter
\eqref{eq:rcritfit}, tied to one of the plasma's IR lengths, numerically
the magnetic length.

If such a transition is real, it would be interesting to discuss its relation to a confinement - deconfinement transition. The reason it might be related at all is that a Wilson loop is the order
parameter of said transition, and the same object whose
commutator we followed obeys an area law on the confined side.  The link is visible in the
saturation value of the correlator: at large times the four-point
function factorises, by large $N$, into
\begin{equation}
c(t)\;\longrightarrow\;
\big|\big\langle W^t_\gamma\big\rangle\big|^2
\big|\big\langle W^{0}_\gamma\big\rangle\big|^2\,,
\label{eq:factorise}
\end{equation}
which in the confined phase is governed by the area law,
\begin{equation}
c(t)\;\sim\;e^{-2\sigma A_\gamma(t)}\,e^{-2\sigma A_\gamma(0)}\,,
\label{eq:arealaw}
\end{equation}
so the value the growth saturates at could by itself be an
order-parameter statement.  Whether the loop then scrambles at a
different rate on the two sides of the transition is not something we can
say: the present computation gives $\lambda_L(R,T,\lambda)$ in the
deconfined phase only, and the confined-phase counterpart is the missing
piece.  We raise the comparison as a motivation, not as a result.

The structure resembles the BFKL resummation
\cite{kuraev_1977,balitsky_1978}.  The large-time limit of an
out-of-time-order correlator is a high-energy limit: the two insertions
sit on opposite sides of the thermal circle and their relative boost
grows like $e^{2\pi t/\beta}$, so the largest eigenvalue should play the
role of a Regge intercept, $j_0-1=\beta\lambda_L/2\pi$
\cite{brower_pomeron_2007}.  The two calculations are similar but not
identical.  BFKL is at zero temperature and its exchanged gluons are
reggeized; ours are dressed by the thermal self-energy instead, which
gives them weight at spacelike momenta and allows the ladder to be placed
on shell.  Our scales are the temperature and the loop size.

Five things are missing from the computation itself.
\begin{enumerate}
\item Equation \eqref{eq:gammacut} is the damping of one gluon with
the IR end of its log cut by hand at $1/2R$, while the true suppression
should be smooth, if anything.
\item Class (i)'s ladder is written down in
\S\ref{sec:bse} and its rung functions in \S\ref{sec:rungs} and
Appendix~\ref{app:classirungs}, but we do not solve it, for the reasons
given in \S\ref{sec:lyap}; whether it grows faster or slower than
class (ii) we cannot say.
\item Like Refs.~\cite{stanford_many-body_2016} and
\cite{steinberg_thermalization_2019}, we worked at zero injected momentum,
and the butterfly velocity \cite{mezei_stanford_2017} requires the
computation we did not do.
\item Adding quarks changes the picture significantly: a cut of the loop becomes a string breaking, with the length scale
$L_*\simeq2m_q/\sigma$ at which pair creation is energetically allowed, and it is the route by which these tools could say something about the
quark--gluon plasma.
\item The same-rail insertions could affect the result rather than
merely extend it.  The end of \S\ref{sec:bse}, with Fig.~\ref{fig:samerail}, records
insertions that connect the two lines of a single commutator, which are the
same order in $\lambda$ and are not excluded by colour, and which
\eqref{eq:finali} does not contain.  If they contribute they enter the
balance between the two terms of the equation, a damping that removes
weight from the rails and a kernel that returns it, and they could enter
either one. 
\end{enumerate}

Four directions beyond it seem worth naming, in rough order of how close
they are to the computation done here.
\begin{enumerate}
\item Grozdanov, Schalm and
Scopelliti \cite{grozdanov_kinetic_2019} showed that for perturbative field
theories the late-time limit of the out-of-time-order correlator is
governed by a Boltzmann-type kinetic equation, their framework suggests a natural setting in which to ask what the present result says about the thermalisation of Yang--Mills, as opposed to the scrambling of one
observable in it.
\item If $\lambda_L(R,T,\lambda)$ is to serve as an order parameter for confinement, the object to study is its behaviour across the transition and its relation to the thermodynamic quantities that are already known to change there.  We have it in the deconfined phase only, and only above the transition temperature; whether it vanishes, jumps, or simply ceases to be defined on the confined side is not something
this computation can say.
\item It would be worth asking whether anything simplifies at the
transition itself: a point where the exponent crosses zero is a natural
place to look for a fixed point, and if one is there the critical loop
size should be calculable.
\item It would be interesting to understand if loop operators are read as states of an effective string rather than as functionals of a gauge field, one could ask for the Hilbert space they span and for the interactions that act on it. The appeal is that a framework of that kind might make the splitting vertex, which here is a contact term needing a regularisation \cite{makeenko_notes_2025}, into something with a controlled definition.
\end{enumerate}

\section*{Acknowledgements}

I am grateful to Sašo Grozdanov for continued support and many discussions, and to Paul Romatschke for his involvement in the early stages of this project. I thank Ofer Aharony, Michael Smolkin, Adam Schwimmer, Koenraad Schalm, Mile Vrbica, Giorgio Frangi, Aviv Barnea, Yaakov Nir Breitstein and Tomer Hadad for discussions and comments.

\appendix

\section{Conventions}
\label{app:conventions}

Colour generators are in the fundamental of $SU(N)$ with
\begin{equation}
\tr\big(T^aT^b\big)=\tfrac12\delta^{ab}\,,\qquad
\tr T^a=0\,,\qquad
\big[T^a,T^b\big]=if^{abc}T^c\,,
\label{eq:appgen}
\end{equation}
and the adjoint matrices are $(F^e)_{ab}=f^{aeb}$, so that
$\tr_{\rm adj}F^e=0$ and $\sum_e\tr\big(F^eF^e\big)=N(N^2-1)$.  The
't~Hooft coupling is $\lambda=g^2N$, held fixed as $N\to\infty$, and every
result below is quoted in $\lambda$.

Momentum integrals and thermal weights are
\begin{equation}
\int_q\equiv\int\!\frac{d^3q}{(2\pi)^3}\,,\qquad
n_B(\omega)=\frac{1}{e^{\beta\omega}-1}\,,\qquad
T=\frac1\beta=1\ \text{in all numerical statements}.
\label{eq:appmeasure}
\end{equation}
The spatial projectors are
\begin{equation}
P^T_{ij}(\hat q)=\delta_{ij}-\hat q_i\hat q_j\,,\qquad
P^L_{\mu\nu}(q)=\frac{q_\mu q_\nu}{q^2}-g_{\mu\nu}-P^T_{\mu\nu}(q)\,,
\label{eq:appproj}
\end{equation}
with $g_{\mu\nu}=\mathrm{diag}(+1,-1,-1,-1)$, the signature in which the
vertices \eqref{eq:G3}--\eqref{eq:G4} and every Lorentz contraction in this
paper are written; the
gauge is general linear covariant, the free propagator being
$P^T+P^L+\xi\,q_\mu q_\nu/(q^2)^2$, and $\xi=0$ throughout, which is Landau
gauge in this parametrisation.

The contour $\gamma$ is a fixed spatial curve $x^i(s)$, taken to be a
circle of radius $R$ in all numerical work, and $R$ appears only in the
dimensionless combination $RT$.

\section{Colour factors of the three contraction classes}
\label{app:colour}

This appendix does in closed form what \S\ref{sec:classes} states: it
counts the powers of $N$ carried by each of the three topologies, and shows
where the cancellation that removes a single exchanged gluon comes from.
The colour factors below are exact in $N$.

Each loop is expanded to two fields, so each of the four traces in
\eqref{eq:master} contributes one factor $\tr(T^aT^b)=\tfrac12\delta^{ab}$.
What distinguishes the classes is which of the four traces are tied to
which.  In class (i) each commutator contracts with itself: the four traces
form two disjoint two-cycles, and summing the adjoint indices gives
\begin{equation}
\mathcal C^{\rm free}_{(i)}
=\Big[\sum_{a,b}\tr\big(T^aT^b\big)^2\Big]^2
=\Big[\tfrac14(N^2-1)\Big]^2
=\frac{(N^2-1)^2}{16}\,.
\label{eq:appfreei}
\end{equation}
In class (ii) the two rails run between the commutators, and the four
traces are chained into a single cycle,
\begin{equation}
\mathcal C^{\rm free}_{(ii)}
=\sum_{a,b,c,e}\tr\big(T^aT^b\big)\tr\big(T^aT^e\big)
\tr\big(T^cT^b\big)\tr\big(T^cT^e\big)
=\frac{N^2-1}{16}\,,
\label{eq:appfreeii}
\end{equation}
one power of $N^2$ down.  Class (iii) chains them in the crossed order and
gives the same $N$ counting as \eqref{eq:appfreeii}.  The ratio
\eqref{eq:appfreei}/\eqref{eq:appfreeii} is $N^2-1$, the colour weight by
which class (ii) sits below class (i) in \S\ref{sec:lyap}.

Attach a single gluon, adjoint index $e$, between two of the four lines and
sum over $e$.  Each factor of the class-(i) skeleton is a closed cycle in
the adjoint, so the insertion of one generator leaves
\begin{equation}
\sum_{\text{lines}}\ \to\ i\,\tr_{\rm adj}F^e
=i\sum_a f^{aea}=0\,,
\label{eq:apptracelemma}
\end{equation}
because $f^{abc}$ is totally antisymmetric and there is no rank-one
invariant tensor in the adjoint.  This is exact in $g$ and in $N$: it is
not a leading-order statement.  The cheapest surviving insertion therefore
carries two gluons, $k=2$, where the same cycle gives
\begin{equation}
\sum_{e_1,e_2}\ i^2\,\tr_{\rm adj}\big(F^{e_1}F^{e_2}\big)
=-N\delta^{e_1e_2}\,,
\label{eq:appk2}
\end{equation}
so a class-(i) rung costs $g^4N^2=\lambda^2$ and the ladder begins at
$O(\lambda^2)$.  On class (ii) the four traces already form one cycle, the
$k=1$ insertion is not a closed adjoint trace, and the same computation
returns $-N$ relative to \eqref{eq:appfreeii}: one gluon survives.

The eight fields of \eqref{eq:master} admit $7!!=105$ perfect pairings, and
most of them are killed before any momentum integral is written.  Each of
the two commutators is a difference of two orderings, so the correlator is
a sum of four operator strings with signs,
\begin{equation}
\big[AA,AA\big]\big[AA,AA\big]^\dagger
=\sum_{e_1,e_2=\pm}e_1e_2\;S_{e_1e_2}\,,
\label{eq:appfourstrings}
\end{equation}
$e=+$ and $e=-$ being the two orderings of that commutator.  A pairing
contributes only if its value distinguishes the four strings.  Take a
pairing that contracts the two fields of the first commutator with each
other at equal time: its Wick value is the same free propagator in all four
strings, because reversing the order of a pair inside a trace at equal time
does not change it, so the four terms enter as $(+1-1-1+1)\times(\text{same
number})=0$.  Sixty-nine of the $105$ die this way, exactly the pairings
whose value is blind to at least one of the two orderings.  The $36$ that
survive are the ones in which every ordering is resolved, and they fall
into the three topologies of \S\ref{sec:classes}: $4$ where each commutator
contracts with itself, $16$ class (ii), $16$ crossed.  The census was done twice, by two
independent enumerations of the $105$ pairings, each evaluating every
pairing in all four orderings on free thermal propagators with the
$\beta/2$ contour shift.

Counting powers of $N$ on the survivors is then the computation above:
\eqref{eq:appfreei} for the $4$, \eqref{eq:appfreeii} for the $16$ of
class (ii), and the same $N$ counting for the crossed $16$.  This is
what makes the two retained topologies enter at different orders in
$1/N^2$ even though they have comparable multiplicities.

The magnitude of every non-vanishing class-(i) placement is the same,
$N^2(N^2-1)/16$ relative to the bare cycle, and only the sign
distinguishes them.  Let $n_s$ count the pairs of cubic vertices that sit
on one and the same rail.  Then
\begin{equation}
\mathcal C=(-1)^{n_s}\,.
\label{eq:coloursign}
\end{equation}
The rule follows from one property of the adjoint generators. Each commutator of the two rail contraction close into a colour cycle, a cubic vertex inserts
one adjoint matrix $F^e$ into it, and a placement's colour factor is the trace of the
product of those matrices in the order the cycle meets them.  The cycle runs out along one rail of the commutator and back along the
other, so it meets one of them against that rail's own direction, and the
generator inserted there enters transposed:
\begin{equation}
\big(F^e\big)^{\!\top}=-F^e\,,\qquad (F^e)_{ab}=f^{aeb}\,,
\label{eq:Ftranspose}
\end{equation}
since $f^{abc}$ is totally antisymmetric.  A pair of vertices sitting on
one and the same rail is therefore either both transposed or neither, and
the two minus signs cancel; a pair split between the two rails picks up
exactly one.  The order in which the cycle meets them does not enter,
since a trace of two matrices is blind to it.  Each of the two cycles of a class-(i) placement
carries exactly one pair, so counting the split pairs and counting the
same-rail pairs differ by a fixed two and give the same sign;
\eqref{eq:coloursign} is written with the latter.  The placements in which one side has
consecutive vertices, families (3) and (6), therefore enter opposite to
the rest.

The colour computation makes this concrete.  By a placement we mean one
labelled way of attaching the rung: which rail each end of each exchanged
gluon lands on, whether the two ends on a side meet at one vertex or two,
and, when both land on the same rail, their order along it.  Enumerating
these gives $144$ of them, and their colour factors, computed as exact
polynomials in $N$, take only three values:
\begin{equation}
\mathcal C^{\rm free}=\frac{(N^2-1)^2}{16}\,,
\qquad
\mathcal C^{\rm placement}=\pm\frac{N^2(N^2-1)}{16}
\ \ \text{or}\ \ 0\,,
\label{eq:threevalues}
\end{equation}
with $44$ of the $144$ vanishing identically, $52$ carrying the plus sign
and $48$ the minus.

Figure~\ref{fig:placement} carries the indices of one placement, and the
contraction is four steps.  Write $(F^{e})_{ab}=f^{aeb}$ for the adjoint
matrices.

\input{worked/placement_figure.tex}

First, a cubic vertex where gluon $e$ meets a rail whose index runs from
$a$ to $a'$ contributes $f^{a e a'}=(F^{e})_{aa'}$, so rail~1 carries
$(F^{e_1})_{aa'}$, rail~2 carries $(F^{e_2})_{bb'}$, and likewise below.
Second, the two loop traces of a commutator contract its two rails with
each other, $\tr(T^{a}T^{b})=\tfrac12\delta^{ab}$ at each end, which is the
statement that rails $1,2$ close into one cycle.  Third, doing those two
sums on the upper cycle,
\begin{align}
\tfrac14\sum_{a,b,a',b'}\delta^{ab}\,\delta^{a'b'}
(F^{e_1})_{aa'}(F^{e_2})_{bb'}
&=\tfrac14\sum_{a,a'}(F^{e_1})_{aa'}(F^{e_2})_{aa'}
\nonumber\\
&=\tfrac14\tr\big(F^{e_1}(F^{e_2})^{\!\top}\big)
=-\tfrac14\tr\big(F^{e_1}F^{e_2}\big),
\label{eq:cycleA}
\end{align}
using $(F^{e})^{\!\top}=-F^{e}$, and the lower cycle gives the same with
$c,d$ in place of $a,b$.  Fourth, the two gluon indices are summed, and with
$\tr(F^{e_1}F^{e_2})=-N\delta^{e_1e_2}$,
\begin{equation}
\mathcal C=\frac{1}{16}\sum_{e_1,e_2}
\big[\tr\big(F^{e_1}F^{e_2}\big)\big]^{2}
=\frac{N^{2}}{16}\sum_{e_1,e_2}\big(\delta^{e_1e_2}\big)^{2}
=\frac{N^{2}(N^{2}-1)}{16}\,,
\label{eq:placementvalue}
\end{equation}
which is the magnitude quoted in \eqref{eq:threevalues}.  Two things are
visible in this that the rail drawing hides.  Each cycle meets exactly two
adjoint matrices, and the trace of two matrices does not depend on their
order, so no placement can be told from another by that order: they all
have this magnitude.  And nothing in \eqref{eq:cycleA} refers to how the
gluon was drawn, which is why a line running across a rail on the page
costs nothing.  What does change between placements is the sign, which
comes from the sense in which each cycle traverses its rails and is the
$(-1)^{n_s}$ of \eqref{eq:coloursign}.  The leading power is $N^4$ in every surviving case, the
same as the free cycle.  That is the answer to the question the drawing
raises: no placement is suppressed by $1/N^2$, however many rails its
gluons appear to cross on the page.  The reason is uniform.  One insertion
puts exactly two exchanged-gluon indices on each of the two cycles, and two
indices on a cycle have no ordering to get wrong, so no placement can be
distinguished from any other by the order in which its cycle meets them.

Ordering does begin to matter once a cycle carries four or more indices,
which is from the second rung of the ladder onwards.  

\section{The kernels on shell}
\label{app:kernels}

This appendix performs the frequency reduction behind
\eqref{eq:onshellBSEi} and \eqref{eq:onshellBSEii}, and evaluates two
representative kernels.  Class (ii) carries the whole
structure in one rail pair, so we write it out; class (i) runs identically,
one rail leg at a time.  Set the caps aside, as in \S\ref{sec:bse}, and write the remaining
pair structure of \eqref{eq:F0boxonshell} times an amplitude
$f(\omega;q)$, equal to one at the free level.
Inserting \eqref{eq:Rbox} into \eqref{eq:BSEi} leaves one exchange
integral,
\begin{equation}
\big[R\,f\big](\omega;k_A^0,q)=-\,\lambda\!\int\!\frac{\dd^4\ell}{(2\pi)^4}\,
\big[\vec n\cdot\mathcal G(\ell)\cdot\vec n\,'\big]\,
f\big(\omega;\,k_A^0-\ell^0,\,q'\big)\,,
\qquad q'=|\vec q-\vec\ell\,|\,,
\label{eq:RFsub}
\end{equation}
with $\vec n=2\vec q$, $\vec n\,'=-2\vec q\,'$ the eikonal factors and the
outer block's delta fixing $k_A^0=s\,\omega_T(q)$.  The delta functions of
the block inside $f$ then perform the $\ell^0$ integral,
\begin{multline}
\int\!\frac{\dd\ell^0}{2\pi}\,
\big[\vec n\cdot\mathcal G(\ell^0,\vec\ell\,)\cdot\vec n\,'\big]\,
2\pi i\Big[\frac{Z_T(q')}{2\omega_T(q')}\Big]^2
\frac{\sum_{s'}\delta\big(s\,\omega_T(q)-\ell^0-s'\omega_T(q')\big)}
{\omega+2i\gamma(q')}\,f(\omega;q')\\
=\,i\Big[\frac{Z_T(q')}{2\omega_T(q')}\Big]^2
\frac{\sum_{s'}\big[\vec n\cdot\mathcal G\big(s\,\omega_T(q)-s'\omega_T(q'),
\vec\ell\,\big)\cdot\vec n\,'\big]}{\omega+2i\gamma(q')}\,f(\omega;q')\,,
\label{eq:l0delta}
\end{multline}
in which the $2\pi$ of the block's prefactor cancels the measure, the
exchange function enters at the frequency the delta fixes, and the residues
of the shifted legs survive as $[Z_T(q')/2\omega_T(q')]^2$.  Of the two
branches, $s'=-s$ puts the exchange near $2\omega_T$, where it has no
spectral weight; the survivor is \eqref{eq:quasielastic}.

For class (i) the same steps must be taken on the zero-momentum block,
where the slaved rails force the two transfers of every two-gluon
insertion to be opposite, $\ell_2=-\ell_1$: the pair scatters as a unit,
and the exchange between the two singlet pairs carries zero total
four-momentum.  Carrying the shifted pair's on-shell structure through
that reduction is not completed in this paper, and we display no class
(i) kernels.  For class (ii), \eqref{eq:l0delta} is already the
kernel of \eqref{eq:finalbox}; the equal-time contractions stay external.

\section{The class (i) rung functions in full}
\label{app:classirungs}

On each side the two gluons attach in one of three ways: at separate
vertices, one on each of that side's two rails; at consecutive vertices
along one rail, with an intermediate rail propagator between them; or
together at one four-gluon vertex.  The combinations give six families,
numbered in Fig.~\ref{fig:rungs} with their placement counts, $64$ labelled placements in all; which families keep an internal loop
can be read off the diagrams.

\begin{figure}[p]
\centering
\begin{tabular}{@{}c@{\hspace{9mm}}l@{}}
\begin{tabular}{@{}c@{}}\rungbox{\raild{}{}{}{}\dKone}\\[2mm]
\rungbox{\raild{}{}{}{}\dKonec}\end{tabular}
& \begin{minipage}{0.40\textwidth}
(1) separate $|$ separate.\\
no internal loop, $\mathcal C=+1$, $4$ placements.\\
Each gluon runs from one line of the first commutator to one line of the
daggered one; each transfer is the shift of the rails it connects.  Both
pairings are drawn: $1\!\to\!3$ with $2\!\to\!4$ above, $1\!\to\!4$ with
$2\!\to\!3$ below.  They carry the same colour factor and differ only in
the momentum routing.
\end{minipage}\\[8mm]
\rungbox{\raild{}{}{}{}\dKtwo}
& \begin{minipage}{0.40\textwidth}
(2) consecutive $|$ consecutive.\\
one internal loop, $\mathcal C=+1$, $16$ placements.\\
Both gluons leave the same rail and land on the same rail; only their sum
is fixed, so one internal momentum is integrated.
\end{minipage}\\[8mm]
\rungbox{\raild{}{}{}{}\dKthree}
& \begin{minipage}{0.40\textwidth}
(3) consecutive $|$ separate, three rails.\\
no internal loop, $\mathcal C=-1$, $16$ placements.\\
The side with consecutive vertices carries an off-shell rail segment between them.
\end{minipage}\\[8mm]
\rungbox{\raild{}{}{}{}\dKfour}
& \begin{minipage}{0.40\textwidth}
(4) quartic $|$ quartic.\\
one internal loop, $\mathcal C=+1$, $4$ placements, symmetry factor $\tfrac12$.\\
Both gluons attach at a single four-gluon vertex on each side.
\end{minipage}\\[8mm]
\rungbox{\raild{}{}{}{}\dKfivea}
& \begin{minipage}{0.40\textwidth}
(5) quartic $|$ separate.\\
no internal loop, $\mathcal C=+1$, $8$ placements (mirror included).
\end{minipage}\\[8mm]
\rungbox{\raild{}{}{}{}\dKfiveb}
& \begin{minipage}{0.40\textwidth}
(6) quartic $|$ consecutive.\\
one internal loop, $\mathcal C=-1$, $16$ placements (mirror included).
\end{minipage}
\end{tabular}
\caption{The seven class-(i) rung families.  Rails
$1,2$ belong to the first commutator and $3,4$ to the daggered one; The families differ in how the two gluons
attach on each side: at separate vertices, at consecutive vertices along one rail, or together at one four-gluon vertex. All count at leading order in $N$, and every family carries the same
colour factor (Appendix~\ref{app:colour}).}
\label{fig:rungs}
\end{figure}

The colour factor of a placement is computed on the loop's colour cycle.  Every non-vanishing placement has the same magnitude,
$N^2(N^2-1)/16$ relative to the bare cycle, and only the sign
distinguishes them; the sign is \eqref{eq:coloursign} of
Appendix~\ref{app:colour}.
The rung functions are written with the explicit powers of $g$ from
their vertices, because each one's colour factor is quoted separately just
above; the two combine to $\lambda^{2}$ for every class-(i) placement, as
$g^4N^2=\lambda^2$, and to $\lambda$ for class (ii).    $\Phi_{n_v}$ is the phase carried by an insertion with $n_v$
vertices, defined below \eqref{eq:G4}; $\varepsilon_j$ and $\varepsilon_j'$ are the
polarisations entering and leaving rail $j$, and
$\vec a\cdot\mathcal G(\ell)\cdot\vec b\equiv a^i\mathcal G_{ij}(\ell)b^j$.
Each rung function is a function of the rail momenta and of its
transfers, and carries no integral of its own; the transfers are
integrated once, in the equation of \S\ref{sec:bse}.  The families whose
diagrams fix only the total transfer, families (2), (4) and (6), keep one internal
loop integral, part of the function itself, as in eq.~(17) of
\cite{stanford_many-body_2016}.

\bigskip

\S\ref{sec:rungs} displays \eqref{eq:R1} and \eqref{eq:R4}.  The
remaining four are written out here.

\bigskip

\noindent One four-gluon vertex and two cubic ones, and no internal
loop.  The four-gluon vertex sits on rail~1 and the two gluons land on
rails 3 and 4:
\filbreak
\begin{center}
\rungwide{\raildmom{k_1'}{k_2'}{k_3'}{k_4'}\dKfivea}
\end{center}
\nopagebreak
\begin{equation}
R^{(5)}\big(\{k\};\ell_1,\ell_2\big)=\Phi_3\,g^4\,
W^{ij}(\varepsilon_1,\varepsilon_1')\,
\mathcal G_{ii'}(\ell_1)\,V^{i'}(\vec q_3,-\vec\ell_1)_{\varepsilon_3\varepsilon_3'}\,
\mathcal G_{jj'}(\ell_2)\,V^{j'}(\vec q_4,-\vec\ell_2)_{\varepsilon_4\varepsilon_4'}\,.
\label{eq:R5}
\end{equation}
Eight placements: the quartic vertex on any of the four rails, against a
split on the other side.

\bigskip

\noindent (2) Consecutive $|$ consecutive ($n_v=4$, $\mathcal C=+1$, one
internal loop).  Both gluons run from rail~1 to rail~3 in the drawn ordering; the
rail pair and the ordering along each rail label distinct placements.  Only
the total transfer $L$ is fixed, so one internal momentum $\ell$ is integrated:
\filbreak
\begin{center}
\rungwide{\raildmom{k_1'}{k_2'}{k_3'}{k_4'}\dKtwo}
\end{center}
\nopagebreak
\begin{align}
R^{(2)}\big(\{k\};L\big)=\Phi_4\,g^4\!
\int\!\frac{\dd^4\ell}{(2\pi)^4}\;
&\Big[V^i(\vec q_1,\vec\ell\,)_{\varepsilon_1\kappa}\,
\Delta^R_T(k_1{-}\ell)\,
V^j(\vec q_1{-}\vec\ell,\vec L{-}\vec\ell\,)_{\kappa\varepsilon_1'}\Big]
\,\mathcal G_{ii'}(\ell)\,\mathcal G_{jj'}(L{-}\ell)
\nonumber\\
\times\;
&\Big[V^{i'}(\vec q_3,-\vec\ell\,)_{\varepsilon_3\kappa'}\,
\Delta^R_T(k_3{+}\ell)\,
V^{j'}(\vec q_3{+}\vec\ell,-\vec L{+}\vec\ell\,)_{\kappa'\varepsilon_3'}\Big]\,,
\label{eq:R2}
\end{align}
where $\kappa,\kappa'$ are summed over the transverse states of the
intermediate segments; the sum over $\kappa,\kappa'$ is each segment's
transverse projector, $\sum_\kappa\varepsilon^i_\kappa\varepsilon^j_\kappa
=P^T_{ij}$, written in components.  A consecutive pair on rail~2 or~4
carries $\widetilde D_T$ in
place of $\Delta^R_T$.  The intermediate segments are not complex
conjugated: a segment is the free contraction of the same pair of branch
operators that produced its rail in $F_0$, so it is the same function
evaluated at the segment momentum.

\bigskip

\noindent (3) Consecutive $|$ separate, three rails
($n_v=4$, $\mathcal C=\boldsymbol{-1}$, no loop).  Two cubic vertices sit on
rail~1 and the two gluons land on different rails below, so the rung
touches three rails and both transfers are fixed:
\filbreak
\begin{center}
\rungwide{\raildmom{k_1'}{k_2'}{k_3'}{k_4'}\dKthree}
\end{center}
\nopagebreak
\begin{align}
R^{(3)}\big(\{k\};\ell_1,\ell_2\big)=-\,\Phi_4\,g^4
&\Big[V^i(\vec q_1,\vec\ell_1)_{\varepsilon_1\kappa}\,
\Delta^R_T(k_1{-}\ell_1)\,
V^j(\vec q_1{-}\vec\ell_1,\vec\ell_2)_{\kappa\varepsilon_1'}\Big]
\nonumber\\
\times\;
&\mathcal G_{ii'}(\ell_1)\,
V^{i'}(\vec q_3,-\vec\ell_1)_{\varepsilon_3\varepsilon_3'}\;
\mathcal G_{jj'}(\ell_2)\,
V^{j'}(\vec q_4,-\vec\ell_2)_{\varepsilon_4\varepsilon_4'}\,.
\label{eq:R3}
\end{align}
Sixteen placements: the consecutive pair on any of the four rails, two orderings each,
against a split on the other side.  These carry the negative colour sign of
\eqref{eq:coloursign}. 

\bigskip

\noindent (6) Quartic $|$ consecutive ($n_v=3$,
$\mathcal C=\boldsymbol{-1}$, one internal loop).  The quartic vertex sits
on rail~1 and both gluons arrive at two consecutive vertices on rail~3, so
only the total transfer $L$ is fixed:
\filbreak
\begin{center}
\rungwide{\raildmom{k_1'}{k_2'}{k_3'}{k_4'}\dKfiveb}
\end{center}
\nopagebreak
\begin{align}
R^{(6)}\big(\{k\};L\big)&=-\,\Phi_3\,g^4\!
\int\!\frac{\dd^4\ell}{(2\pi)^4}\,
W^{ij}(\varepsilon_1,\varepsilon_1')\,
\mathcal G_{ii'}(\ell)\,\mathcal G_{jj'}(L{-}\ell)
\nonumber\\
&\hspace{16mm}\times
\Big[V^{i'}(\vec q_3,-\vec\ell\,)_{\varepsilon_3\kappa'}\,
\Delta^R_T(k_3{+}\ell)\,
V^{j'}(\vec q_3{+}\vec\ell,-\vec L{+}\vec\ell\,)_{\kappa'\varepsilon_3'}\Big]\,.
\label{eq:R6}
\end{align}
The mirror placements, with the quartic vertex on the lower side, are
counted in the multiplicities of Fig.~\ref{fig:rungs}.

One normalisation is recorded here for completeness.  Each
class-(i) insertion is built with the factor $\lambda^{2}/8$, with a
further $\tfrac12$ on the quartic$|$quartic placement, and each placement
then enters with the colour sign of \eqref{eq:coloursign} and the
multiplicity listed with \eqref{eq:finali}.  The $\lambda^{2}/8$ is
$g^{4}$ times the colour factor an added rung supplies, which is $N^{2}/8$
at large $N$.

\section{Numerical procedure}
\label{app:numerics}

The eigenvalue problem \eqref{eq:finalbox} is solved by  diagonalisation. The rail momentum lies on a geometric grid $q\in[0.5,4]\,T$, hard by
construction, with $N_q=48$ points; class (ii) has one momentum to carry,
so its matrix has dimension $48$.  Thermal functions
$\omega_T,Z_T,\gamma$ are evaluated on
$[10^{-3},12]\,T$; the spectral functions of the exchanged lines, in the
convention $\rho=2\,\mathrm{Im}\,\Delta^R$ of \eqref{eq:raillines}, on a grid of
$481\times241$ in the frequency ratio and the momentum.  A shifted
leg that leaves the rail grid is masked away, and the fraction of transfer
weight lost to that mask is printed by every run.

The transfer momentum of a rung runs over
$|\ell|\in[\sqrt{\lambda}\,m_D,\,2\,m_D]$: the upper end is where the
spectral weight of a soft dressed line dies away, the lower end is the
magnetic-scale estimate at which every integral over soft exchange stops
being perturbative, and with $m_D=0.183\,T$ at $\lambda=0.1$.  The
transfer frequency is fixed quasi-elastically by
\eqref{eq:quasielastic}, and since the group velocity of a transverse mode
is below one, every exchange is spacelike, on the Landau cut; at $\max|\ell^0|/|\vec\ell\,|=0.94$--$0.98$.  The angle between a
transfer and the rail carries a narrow feature near the perpendicular
direction, of width set by the Landau-damping denominator, and the angular
grid uses a finer central panel of half-width
$\min(0.4,\,40\lambda/\pi)$.  Shifted arguments of the eigenfunction
are interpolated linearly in $\ln q$.

$\lambda_L$ is evaluated
on seventy geometric points in $2R\,m_D\in[0.5,16]$, and $2R^*m_D$ is the linear interpolation of the last change of sign from growth to decay.  The error of that location is a fraction of the spacing and grows where the exponent crosses shallowly.

The matrix is built in full and all of its
eigenvalues are computed at once, with \texttt{numpy}; $\lambda_L$ is the
largest real part.

\bibliography{references}
\bibliographystyle{unsrt}

\end{document}

%% file: worked/worldsheet_figure.tex
\begin{figure}[htbp]
\centering
\begin{tikzpicture}[scale=0.8,>=stealth,thick,every node/.append style={font=\footnotesize}]
  \draw[->,gray,thin] (0,-3.4) -- (9.0,-3.4) node[right,black]{$t$};
  \foreach \tx/\lab in {2.2/{$1\to2$},4.7/{$2\to3$},6.6/{$3\to4$}}{
    \draw[gray,thin] (\tx,-3.3) -- (\tx,-3.5);
    \node[gray] at (\tx,-3.75) {\lab};
  }
  \draw (0,0.55) -- (1.5,0.55) .. controls (2.1,0.7) and (2.5,1.3) .. (2.9,1.9);
  \draw (0,-0.55) -- (1.5,-0.55) .. controls (2.1,-0.7) and (2.5,-1.3) .. (2.9,-1.9);
  \draw (0,0) ellipse (0.14 and 0.55);
  \node at (-0.5,0) {$\gamma$};
  \draw (2.2,0) .. controls (2.45,0.45) and (2.65,0.75) .. (2.9,0.95);
  \draw (2.2,0) .. controls (2.45,-0.45) and (2.65,-0.75) .. (2.9,-0.95);
  \filldraw[red] (2.2,0) circle (1.8pt);
  \draw (2.9,1.9) -- (4.0,1.9);   \draw (2.9,0.95) -- (4.0,0.95);
  \draw (2.9,-0.95) -- (6.0,-0.95);  \draw (2.9,-1.9) -- (6.0,-1.9);
  \draw (4.0,1.9) .. controls (4.7,2.15) and (5.2,2.4) .. (5.8,2.5) -- (8.2,2.5);
  \draw (4.0,0.95) .. controls (4.7,0.85) and (5.2,0.75) .. (5.8,0.7) -- (8.2,0.7);
  \draw (4.7,1.42) .. controls (5.1,1.7) and (5.4,1.85) .. (5.8,1.9) -- (8.2,1.9);
  \draw (4.7,1.42) .. controls (5.1,1.34) and (5.4,1.3) .. (5.8,1.3) -- (8.2,1.3);
  \filldraw[red] (4.7,1.42) circle (1.6pt);
  \draw (6.0,-0.95) .. controls (6.6,-0.85) and (7.1,-0.75) .. (7.6,-0.7) -- (8.2,-0.7);
  \draw (6.0,-1.9) .. controls (6.6,-2.15) and (7.1,-2.4) .. (7.6,-2.5) -- (8.2,-2.5);
  \draw (6.6,-1.42) .. controls (7.0,-1.34) and (7.3,-1.3) .. (7.6,-1.3) -- (8.2,-1.3);
  \draw (6.6,-1.42) .. controls (7.0,-1.7) and (7.3,-1.85) .. (7.6,-1.9) -- (8.2,-1.9);
  \filldraw[red] (6.6,-1.42) circle (1.6pt);
  \foreach \cy in {2.2,1.0,-1.0,-2.2}{ \draw (8.2,\cy) ellipse (0.12 and 0.3); }
  \node[align=center] at (9.15,0) {4 loops\\ at $t$};
\end{tikzpicture}
\caption{A single connected history drawn as a worldsheet (time runs to the
right).  One tube enters and three pairs of pants split it into four loops,
one split at a time, $s=1\to2\to3\to4$: red dots mark the cuts, the ticks
their times.  A tube not being cut continues as a cylinder, so the surface
stays connected.}
\label{fig:worldsheet}
\end{figure}

%% file: worked/contour_figure.tex
\begin{figure}[htbp]
\centering
\resizebox{0.98\textwidth}{!}{%
\begin{tikzpicture}[scale=1.0,>=stealth]

\newcommand{\contour}[2]{%
\begin{scope}[xshift=#1cm]
  \draw[thick] (0,1.15) -- (0,0.85);
  \draw[thick] (0,0.85) -- (0,-0.85);
  \draw[thick] (0,-0.85) -- (0,-1.15);
  \node[font=\scriptsize,left] at (-0.08,0) {$\tfrac{\beta}{2}$};
  \draw[thick] (0,1.15) -- (3.55,1.15)
        .. controls (3.95,1.15) and (3.95,0.85) .. (3.55,0.85) -- (0,0.85);
  \draw[thick] (0,-0.85) -- (3.55,-0.85)
        .. controls (3.95,-0.85) and (3.95,-1.15) .. (3.55,-1.15) -- (0,-1.15);
  \node[font=\scriptsize] at (4.25,1.0) {$t$};
  \node[font=\scriptsize] at (4.25,-1.0) {$t$};
  \node[font=\scriptsize] at (-0.28,1.0) {$0$};
  \node[font=\scriptsize] at (-0.28,-1.0) {$0$};
  \node[font=\small] at (2.0,-2.15) {#2};
  \foreach \x in {3.05,3.45} { \filldraw (\x,1.15) circle (1.5pt);
                               \filldraw (\x,-0.85) circle (1.5pt); }
  \foreach \x in {0.45,0.85} { \filldraw (\x,0.85) circle (1.5pt);
                               \filldraw (\x,-1.15) circle (1.5pt); }
\end{scope}}

\contour{0}{class (i): four rails}
\begin{scope}[xshift=0cm]
  \draw[thick,blue] (3.05,1.15) .. controls (2.2,1.55) and (1.3,1.55) .. (0.45,0.85);
  \draw[thick,blue] (3.45,1.15) .. controls (2.4,1.75) and (1.4,1.75) .. (0.85,0.85);
  \draw[thick,blue] (3.05,-0.85) .. controls (2.2,-1.55) and (1.3,-1.55) .. (0.45,-1.15);
  \draw[thick,blue] (3.45,-0.85) .. controls (2.4,-1.75) and (1.4,-1.75) .. (0.85,-1.15);
\end{scope}

\contour{6.6}{class (ii): two rails, two caps}
\begin{scope}[xshift=6.6cm]
  \draw[thick,blue] (3.05,1.15) .. controls (2.2,1.55) and (1.3,1.55) .. (0.45,0.85);
  \draw[thick,blue] (3.05,-0.85) .. controls (2.2,-1.55) and (1.3,-1.55) .. (0.45,-1.15);
  \draw[thick,red] (3.45,1.15) .. controls (4.35,0.6) and (4.35,-0.3) .. (3.45,-0.85);
  \draw[thick,red] (0.85,0.85) .. controls (1.65,0.3) and (1.65,-0.6) .. (0.85,-1.15);
  \node[font=\scriptsize,red] at (4.75,0.15) {cap};
  \node[font=\scriptsize,red] at (2.05,-0.15) {cap};
\end{scope}

\contour{13.2}{class (iii): crossed (out of scope)}
\begin{scope}[xshift=13.2cm]
  \draw[thick,blue] (3.05,1.15) .. controls (2.2,1.55) and (1.3,1.55) .. (0.45,0.85);
  \draw[thick,blue] (3.05,-0.85) .. controls (2.2,-1.55) and (1.3,-1.55) .. (0.45,-1.15);
  \draw[thick,blue] (3.45,1.15) .. controls (2.9,0.25) and (1.9,-0.45) .. (0.85,-1.15);
  \draw[thick,blue] (3.45,-0.85) .. controls (2.9,-0.05) and (1.9,0.55) .. (0.85,0.85);
\end{scope}

\end{tikzpicture}}
\caption{The eight fields of \eqref{eq:master} on the out-of-time-order
contour, and the three ways of contracting them.  Each horizontal fold is
one real-time evolution out to time $t$ and back, carrying one commutator;
the two folds are separated by half the thermal circle, which is what puts
the offset $\beta/2$ on any propagator running between them.  Filled dots
are the field insertions, four at time $t$ and four at time $0$.  In class
(i) each commutator contracts with itself, giving four lines, the rails.  In class (ii) one line per commutator spans
the interval and the remaining fields contract across the folds at equal
time, forming the two caps, that have no time extent. In class (iii) one gluon propagates within each fold and the remaining
pair crosses between them at unequal times, so the skeleton on which a
ladder would be iterated is not fixed; it is not treated here.  All lines in this and the following figures are
gluon propagators: rails are drawn straight and exchanged gluons wavy.}
\label{fig:contour}
\end{figure}

%% file: worked/pinch_contour_figure.tex
\begin{figure}[htbp]
\centering
\begin{tikzpicture}[scale=1.0,>=stealth,
  every node/.append style={font=\footnotesize}]

\draw[->,thick] (-4.9,0) -- (5.3,0) node[right] {$\mathrm{Re}\,k_1^{0}$};
\draw[->,thick] (0,-1.75) -- (0,1.95) node[above] {$\mathrm{Im}\,k_1^{0}$};

\draw[very thick] (-4.6,0) -- (5.0,0);
\draw[very thick,->] (-1.4,0) -- (-1.0,0);

\foreach \x in {-3.3, 2.6}{\filldraw[blue!70!black] (\x,-0.62) circle (2.7pt);}
\node[blue!70!black,below right] at (-3.35,-0.74) {$-\omega_T(q_1)-i\gamma_1$};
\node[blue!70!black,left] at (2.40,-0.62) {$+\omega_T(q_1)-i\gamma_1$};

\foreach \x in {-1.5, 2.6}{\filldraw[red!75!black] (\x,0.62) circle (2.7pt);}
\node[red!75!black,above] at (-1.5,0.78) {$\omega+\omega_T(q_1)-2\omega_T(q_3)+i\gamma_3$};
\node[red!75!black,above] at (2.6,0.78) {$\omega+\omega_T(q_1)+i\gamma_3$};

\draw[<->,very thick,green!45!black] (2.6,0.44) -- (2.6,-0.44);
\node[green!45!black,above left,inner sep=1pt] at (2.5,0.06)
  {$\omega+i(\gamma_1+\gamma_3)$};
\node[green!45!black,align=left,anchor=north] at (2.6,-1.15)
  {the two are $\omega+i(\gamma_1{+}\gamma_3)$ apart};

\end{tikzpicture}
\caption{The $k_1^{0}$ contour of \eqref{eq:R13}, in the sector
\eqref{eq:pinch} that survives.  Rail~1 enters as
$\Delta^R_T(k_1^0)$, so its two poles sit $\gamma(q_1)$ below the real
axis, in blue; rail~3 enters at the frequency the constraint leaves it, which puts
its poles $\gamma(q_3)$ above, in red.  The contour runs along the real
axis and is closed below, so it picks up rail~1's poles.  When the two poles marked in green approach
each other the contour has to pass between them, and the denominator of
\eqref{eq:R13} is then no larger than $\gamma(q_1)+\gamma(q_3)$.}
\label{fig:pinch}
\end{figure}

%% file: worked/ladder_on_loops_figure.tex
\begin{figure}[htbp]
\centering
\resizebox{0.72\textwidth}{!}{%
\begin{tikzpicture}[scale=1.0,>=stealth,
  rail/.style={decorate,decoration={snake,amplitude=1.2pt,segment length=5.5pt},
               thick,blue!70!black},
  rung/.style={decorate,decoration={snake,amplitude=1.2pt,segment length=5.5pt},
               thick,red!75!black}]

\draw [black, ultra thick] (0,2) to [bend left=40] (0,0.7);
\draw [black, ultra thick] (0,2) to [bend left=-40] (0,0.7);
\draw [black, ultra thick] (0,-2) to [bend left=40] (0,-0.7);
\draw [black, ultra thick] (0,-2) to [bend left=-40] (0,-0.7);
\draw [black, ultra thick] (0,-0.7) to [bend left=-50] (1,0);
\draw [black, ultra thick] (0,0.7) to [bend left=50] (1,0);
\draw [black, ultra thick] (0,2) to [bend left=20] (1.5,1.6);
\draw [black, ultra thick] (1.5,1.6) to [bend left=-30] (4.5,1.6);
\draw [black, ultra thick] (0,-2) to [bend left=-20] (1.5,-1.6);
\draw [black, ultra thick] (1.5,-1.6) to [bend left=30] (4.5,-1.6);
\draw [black, ultra thick] (4.5,1.6) to [bend left=20] (6,2);
\draw [black, ultra thick] (4.5,-1.6) to [bend left=-20] (6,-2);
\draw [black, ultra thick] (6,2) to [bend left=40] (6,0.7);
\draw [black, ultra thick] (6,2) to [bend left=-40] (6,0.7);
\draw [black, ultra thick] (6,-2) to [bend left=40] (6,-0.7);
\draw [black, ultra thick] (6,-2) to [bend left=-40] (6,-0.7);
\draw [black, ultra thick] (6,-0.7) to [bend left=50] (5,0);
\draw [black, ultra thick] (6,0.7) to [bend left=-50] (5,0);

\draw[rail] (0.22,1.78) .. controls (1.57,0.740) and (4.43,0.740) .. (5.78,1.78);
\draw[rail] (0.34,1.12) .. controls (1.69,0.373) and (4.31,0.373) .. (5.66,1.12);
\draw[rail] (0.34,-1.12) .. controls (1.69,-0.373) and (4.31,-0.373) .. (5.66,-1.12);
\draw[rail] (0.22,-1.78) .. controls (1.57,-0.740) and (4.43,-0.740) .. (5.78,-1.78);

\draw[rung] (1.55,1.174) -- (1.55,-0.700);
\filldraw[green!45!black] (1.55,1.174) circle (2.0pt);
\filldraw[green!45!black] (1.55,-0.700) circle (2.0pt);
\draw[rung] (2.35,0.587) -- (2.35,-1.033);
\filldraw[green!45!black] (2.35,0.587) circle (2.0pt);
\filldraw[green!45!black] (2.35,-1.033) circle (2.0pt);
\draw[rung] (3.15,1.002) -- (3.15,-1.002);
\filldraw[green!45!black] (3.15,1.002) circle (2.0pt);
\filldraw[green!45!black] (3.15,-1.002) circle (2.0pt);
\draw[rung] (3.95,0.618) -- (3.95,-0.618);
\filldraw[green!45!black] (3.95,0.618) circle (2.0pt);
\filldraw[green!45!black] (3.95,-0.618) circle (2.0pt);
\draw[rung] (4.65,1.229) -- (4.65,-0.745);
\filldraw[green!45!black] (4.65,1.229) circle (2.0pt);
\filldraw[green!45!black] (4.65,-0.745) circle (2.0pt);

\node[blue!70!black,font=\small,anchor=south] at (3.0,2.21) {rails};
\draw[blue!70!black,thin] (3.0,2.16) -- (2.20,1.42);
\node[red!75!black,font=\small,anchor=west] at (5.90,0.30) {exchanges};
\draw[red!75!black,thin] (5.85,0.30) -- (4.72,0.30);

\node[font=\small] at (-1.15,1.35) {$W^{0\dagger}_{\gamma}$};
\node[font=\small] at (-1.15,-1.35) {$W^{0\dagger}_{\gamma}$};
\node[font=\small] at (7.15,1.35) {$W^{t}_{\gamma}$};
\node[font=\small] at (7.15,-1.35) {$W^{t}_{\gamma}$};

\filldraw (0,1.35) circle (1.4pt);
\draw[<->,thick] (0,1.35) -- (0,1.97);
\node[font=\small] at (-0.46,1.68) {$R$};

\end{tikzpicture}}
\caption{The class-(i) ladder on the worldsheet it lives on.  The surface is
the leading term of Fig.~\ref{fig:genusexp}
Each commutator contributes two gluon lines running the length of the
surface, the rails, and the insertions that build the ladder are the
gluons exchanged between the two commutators, drawn vertically with their
vertices marked.  The correlator grows or decays according to whether these
exchanges outweigh the damping of the rails, and the loop radius $R$ enters
because an exchange cannot connect points of the contour separated by more
than $2R$ (Fig.~\ref{fig:cutoffgeom}).}
\label{fig:ladderloops}
\end{figure}

%% file: worked/samerail_figure.tex
\begin{figure}[htbp]
\centering
\begin{tikzpicture}[xscale=1.05,yscale=0.70,>=stealth,
  every node/.append style={font=\footnotesize}]

\begin{scope}
  \foreach \y/\l in {3/1, 2.2/2, 0.8/3, 0/4}{
    \draw[thick] (0,\y)--(6.2,\y); \node[left] at (0,\y) {$\l$};}
  \foreach \x in {1.3,3.1,4.9}{
    \draw[gl,red!75!black] (\x,3)--(\x,0.8);
    \draw[gl,red!75!black] (\x+0.55,2.2)--(\x+0.55,0);}
  \node at (3.1,-1.15) {between the two commutators};
\end{scope}

\begin{scope}[shift={(8.4,0)}]
  \foreach \y/\l in {3/1, 2.2/2, 0.8/3, 0/4}{
    \draw[thick] (0,\y)--(6.2,\y); \node[left] at (0,\y) {$\l$};}
  \foreach \x in {1.0,2.8,4.6}{
    \draw[gl,blue!70!black] (\x,3)--(\x,2.2);}
  \foreach \x in {1.9,3.7,5.5}{
    \draw[gl,blue!70!black] (\x,0.8)--(\x,0);}
  \node at (3.1,-1.15) {within one commutator};
\end{scope}

\end{tikzpicture}
\caption{Two kinds of insertion on the same four rails.  Left, in red, the
rungs summed in this paper: they join a line of the first commutator to a
line of the daggered one, and it is their iteration that produces
\eqref{eq:finali}.  Right, in blue, insertions that connect the two lines
of a single commutator.  These are the same order in $\lambda$ and are not
excluded by colour, but they have not been computed and \eqref{eq:finali}
does not contain them.  Whether they contribute at all, and if so how they
divide between the damping and the kernel, is left open.}
\label{fig:samerail}
\end{figure}

%% file: worked/cutoff_geometry_figure.tex
\begin{figure}[htbp]
\centering
\resizebox{0.46\textwidth}{!}{%
\begin{tikzpicture}[scale=1.0,>=stealth,
  gluon/.style={decorate,decoration={coil,amplitude=3.2pt,segment length=5.2pt},
                thick,green!55!black}]

\draw[->,thick,blue!70!black] (0,-2.35) -- (0,3.45);
\draw[->,thick,blue!70!black] (0,-1.1) -- (-3.5,-2.75);
\draw[->,thick,blue!70!black] (0,-1.1) -- (4.1,-2.15);

\draw[very thick,blue!70!black] (0,2.0) ellipse (2.85 and 0.85);
\node[blue!70!black] at (2.35,1.28) {$\gamma$};
\filldraw[blue!70!black] (2.05,1.42) circle (2.2pt);
\node[blue!70!black,font=\small] at (3.35,1.10) {$x(s'_{2})$};

\draw[very thick,blue!70!black] (0,-1.1) ellipse (2.85 and 0.85);
\node[blue!70!black] at (0.85,-1.95) {$\gamma$};
\filldraw[blue!70!black] (-2.35,-1.52) circle (2.2pt);
\node[blue!70!black,font=\small] at (-3.35,-1.22) {$x(s'_{1})$};

\draw[gluon] (-2.35,-1.52) .. controls (-1.3,-0.35) and (0.9,0.5) .. (2.05,1.42);

\node at (0,-3.55) {$\big|\,x(s'_{1})-x(s'_{2})\,\big|\;\le\;2R$};

\end{tikzpicture}}
\caption{The geometric origin of the IR cutoff.  The same spatial
contour is drawn at the two times entering the correlator; a gluon exchanged
in the course of the evolution runs between two points of the contour, so
its endpoints can never be separated by more than the diameter, $2R$.
A mode of momentum $\ell\lesssim1/2R$ cannot tell the two lines of a
commutator apart: it couples to their total colour charge, which is zero,
through the factor $4\sin^{2}(\vec\ell\cdot\vec d/2)$, so it does not
damp them.  This is what puts $1/2R$ at the IR end of the log
in \eqref{eq:gammacut}, and it is why the exponent depends on the size of
the loop.}
\label{fig:cutoffgeom}
\end{figure}

%% file: worked/cobordisms_figure.tex
\begin{figure}[htbp]
\centering
\begin{tikzpicture}[scale=0.92]
  \draw[->] (-1.5,0) -- (-1.5,3.6);
  \node[left] at (-1.5,3.4) {\footnotesize $t$};
  \begin{scope}
  \draw (0,0) ellipse (0.55 and 0.16);
  \draw (-0.55,0) -- (-0.55,3.6);
  \draw (0.55,0) -- (0.55,3.6);
  \draw (-0.55,3.6) arc (180:360:0.55 and 0.16);
  \draw[dashed] (0.55,3.6) arc (0:180:0.55 and 0.16);
  \node at (0,-0.5) {\footnotesize $W^{0}_{\gamma}$};
  \node at (0,4.1) {\footnotesize $W^{t}_{\gamma}$};
  \node at (0,-1.05) {\footnotesize (a) $g=0$};
  \end{scope}
  \begin{scope}[shift={(4.3,0)}]
  \draw (0,0) ellipse (0.55 and 0.16);
  \draw (-0.55,0) -- (-0.55,3.6);
  \draw (0.55,0) -- (0.55,3.6);
  \draw (-0.55,3.6) arc (180:360:0.55 and 0.16);
  \draw[dashed] (0.55,3.6) arc (0:180:0.55 and 0.16);
  \draw (0,1.2) .. controls (-0.26,1.55) and (-0.26,2.05) .. (0,2.4);
  \draw (0,1.2) .. controls (0.26,1.55) and (0.26,2.05) .. (0,2.4);
  \filldraw (0,1.2) circle (1.3pt);
  \filldraw (0,2.4) circle (1.3pt);
  \node[right] at (0.58,1.1) {\scriptsize split $t_{1}$};
  \node[right] at (0.58,2.5) {\scriptsize join $t_{2}$};
  \node at (0,-1.05) {\footnotesize (b) $g=1$};
  \end{scope}
\end{tikzpicture}
\caption{Surfaces spanned between the two loop insertions of the
connected two-point function \eqref{eq:cobordism}.  Time runs upward, from
$W^{0}_{\gamma}$ to $W^{t}_{\gamma}$.  (a) Genus zero is the cylinder, the
loop propagating without splitting.  (b) Genus one: a split at $t_{1}$ and
a join at $t_{2}$ bound a handle, which costs $1/N^{2}$ from the topology
and $\varepsilon^{2}$ from its two cut events.  Each further handle repeats
that pattern.}
\label{fig:cobordisms}
\end{figure}

%% file: worked/genus_expansion_figure.tex
\begin{figure}[htbp]
\centering
\resizebox{0.98\textwidth}{!}{%
\begin{tikzpicture}[scale=0.52]
\begin{scope}[xshift=10cm]
\draw [black, ultra thick] (0,2) to [bend left=40] (0,0.7);
\draw [black, ultra thick] (0,2) to [bend left=-40] (0,0.7);
\draw [black, ultra thick] (0,-2) to [bend left=40] (0,-0.7);
\draw [black, ultra thick] (0,-2) to [bend left=-40] (0,-0.7);
\draw [black, ultra thick] (0,-0.7) to [bend left=-50] (1,0);
\draw [black, ultra thick] (0,0.7) to [bend left=50] (1,0);
\draw [black, ultra thick] (0,2) to [bend left=20] (1.5,1.6);
\draw [black, ultra thick] (1.5,1.6) to [bend left=-30] (4.5,1.6);
\draw [black, ultra thick] (0,-2) to [bend left=-20] (1.5,-1.6);
\draw [black, ultra thick] (1.5,-1.6) to [bend left=30] (4.5,-1.6);
\draw [black, ultra thick] (4.5,1.6) to [bend left=20] (6,2);
\draw [black, ultra thick] (4.5,-1.6) to [bend left=-20] (6,-2);
\draw [black, ultra thick] (6,2) to [bend left=40] (6,0.7);
\draw [black, ultra thick] (6,2) to [bend left=-40] (6,0.7);
\draw [black, ultra thick] (6,-2) to [bend left=40] (6,-0.7);
\draw [black, ultra thick] (6,-2) to [bend left=-40] (6,-0.7);
\draw [black, ultra thick] (6,-0.7) to [bend left=50] (5,0);
\draw [black, ultra thick] (6,0.7) to [bend left=-50] (5,0);
\draw [black, ultra thick] (10,2) to [bend left=40] (10,0.7);
\draw [black, ultra thick] (10,2) to [bend left=-40] (10,0.7);
\draw [black, ultra thick] (10,-2) to [bend left=40] (10,-0.7);
\draw [black, ultra thick] (10,-2) to [bend left=-40] (10,-0.7);
\draw [black, ultra thick] (10,-0.7) to [bend left=-50] (11,0);
\draw [black, ultra thick] (10,0.7) to [bend left=50] (11,0);
\draw [black, ultra thick] (10,2) to [bend left=20] (11.5,1.6);
\draw [black, ultra thick] (11.5,1.6) to [bend left=-30] (14.5,1.6);
\draw [black, ultra thick] (10,-2) to [bend left=-20] (11.5,-1.6);
\draw [black, ultra thick] (11.5,-1.6) to [bend left=30] (14.5,-1.6);
\draw [black, ultra thick] (14.5,1.6) to [bend left=20] (16,2);
\draw [black, ultra thick] (14.5,-1.6) to [bend left=-20] (16,-2);
\draw [black, ultra thick] (16,2) to [bend left=40] (16,0.7);
\draw [black, ultra thick] (16,2) to [bend left=-40] (16,0.7);
\draw [black, ultra thick] (16,-2) to [bend left=40] (16,-0.7);
\draw [black, ultra thick] (16,-2) to [bend left=-40] (16,-0.7);
\draw [black, ultra thick] (16,-0.7) to [bend left=50] (15,0);
\draw [black, ultra thick] (16,0.7) to [bend left=-50] (15,0);
\draw [black, ultra thick] (12, 0.2) to [bend left=-50] (14, 0.2);
\draw [black, ultra thick] (12.3,0) to [bend left=50] (13.7,0);
\draw [black, ultra thick] (20,2) to [bend left=40] (20,0.7);
\draw [black, ultra thick] (20,2) to [bend left=-40] (20,0.7);
\draw [black, ultra thick] (20,-2) to [bend left=40] (20,-0.7);
\draw [black, ultra thick] (20,-2) to [bend left=-40] (20,-0.7);
\draw [black, ultra thick] (20,-0.7) to [bend left=-50] (21,0);
\draw [black, ultra thick] (20,0.7) to [bend left=50] (21,0);
\draw [black, ultra thick] (20,2) to [bend left=20] (21.5,1.6);
\draw [black, ultra thick] (21.5,1.6) to [bend left=-30] (24.5,1.6);
\draw [black, ultra thick] (20,-2) to [bend left=-20] (21.5,-1.6);
\draw [black, ultra thick] (21.5,-1.6) to [bend left=30] (24.5,-1.6);
\draw [black, ultra thick] (24.5,1.6) to [bend left=20] (26,2);
\draw [black, ultra thick] (24.5,-1.6) to [bend left=-20] (26,-2);
\draw [black, ultra thick] (26,2) to [bend left=40] (26,0.7);
\draw [black, ultra thick] (26,2) to [bend left=-40] (26,0.7);
\draw [black, ultra thick] (26,-2) to [bend left=40] (26,-0.7);
\draw [black, ultra thick] (26,-2) to [bend left=-40] (26,-0.7);
\draw [black, ultra thick] (26,-0.7) to [bend left=50] (25,0);
\draw [black, ultra thick] (26,0.7) to [bend left=-50] (25,0);
\draw [black, ultra thick] (21.4, 0.2) to [bend left=-50] (22.6, 0.2);
\draw [black, ultra thick] (21.6,0) to [bend left=50] (22.4,0);
\draw [black, ultra thick] (23.2, 0.2) to [bend left=-50] (24.4, 0.2);
\draw [black, ultra thick] (23.4,0) to [bend left=50] (24.2,0);
\draw (8, 0) node{$+ \;\; N^{-2}\varepsilon^{2}$};
\draw (18, 0) node{$+ \;\; N^{-4}\varepsilon^{4}$};
\draw (28.4, 0) node{$+ \cdots$};
\draw (0,2.9) node{\small $W^{0\dagger}_{\gamma}$};
\draw (0,-2.9) node{\small $W^{0\dagger}_{\gamma}$};
\draw (6,2.9) node{\small $W^{t}_{\gamma}$};
\draw (6,-2.9) node{\small $W^{t}_{\gamma}$};
\end{scope}
\end{tikzpicture}}
\caption{Schematic representation of the proposed expansion of the loop
out-of-time-order correlator: a genus expansion of $2\to2$ scattering of
closed flux tubes, two loops entering at time $0$ and two leaving at time
$t$.  It is the same topological series as the perturbative string
expansion, where each handle carries $g_s^{2}$; here the weight of a handle
is fixed by the two gradings of Sec.~\ref{sec:cobordism}, one factor
$N^{-2}$ from the 't~Hooft counting and one factor $\varepsilon^{2}$ from
the two cut events, one splitting and one joining, that a handle
contains, so $g_s^{2}\to N^{-2}\varepsilon^{2}$.  The leading term is the object a Lyapunov exponent would be read from; as
discussed in the text, the proposal identifies that object but no
controlled amplitude for it is available in a confining background.}
\label{fig:genusexp}
\end{figure}

%% file: worked/placement_figure.tex
\begin{figure}[htbp]
\centering
\begin{tikzpicture}[scale=1.0,>=stealth,
  every node/.append style={font=\footnotesize}]

\foreach \y/\l in {3.0/1, 2.1/2, 0.9/3, 0/4}{
  \draw[thick] (0,\y)--(7.0,\y);
  \node[left] at (-0.75,\y) {\scriptsize $\l$};}

\filldraw (2.6,3.0) circle (1.7pt);  \filldraw (2.6,0.9) circle (1.7pt);
\filldraw (4.6,2.1) circle (1.7pt);  \filldraw (4.6,0)   circle (1.7pt);

\draw[gl,red!75!black] (2.6,3.0)--(2.6,0.9);
\draw[gl,red!75!black] (4.6,2.1)--(4.6,0);
\node[red!75!black,left]  at (2.42,1.50) {$e_1$};
\node[red!75!black,right] at (4.78,1.50) {$e_2$};

\node[above] at (1.3,3.03) {$a$};        \node[above] at (5.0,3.03) {$a'$};
\node[above] at (2.1,2.13) {$b$};        \node[above] at (5.9,2.13) {$b'$};
\node[above] at (1.3,0.93) {$c$};        \node[above] at (5.0,0.93) {$c'$};
\node[above] at (2.1,0.03) {$d$};        \node[above] at (5.9,0.03) {$d'$};

\draw[thick,blue!70!black] (0,3.0) .. controls (-0.5,2.7) and (-0.5,2.4) .. (0,2.1);
\draw[thick,blue!70!black] (7.0,3.0) .. controls (7.5,2.7) and (7.5,2.4) .. (7.0,2.1);
\draw[thick,blue!70!black] (0,0.9) .. controls (-0.5,0.6) and (-0.5,0.3) .. (0,0);
\draw[thick,blue!70!black] (7.0,0.9) .. controls (7.5,0.6) and (7.5,0.3) .. (7.0,0);
\node[blue!70!black] at (-1.35,2.55) {$\tfrac12\delta^{ab}$};
\node[blue!70!black] at (8.35,2.55) {$\tfrac12\delta^{a'b'}$};
\node[blue!70!black] at (-1.35,0.45) {$\tfrac12\delta^{cd}$};
\node[blue!70!black] at (8.35,0.45) {$\tfrac12\delta^{c'd'}$};

\end{tikzpicture}
\caption{The colour indices of one placement, and what closes them.  Each
rail is an adjoint line and each exchanged gluon carries its own adjoint
index, $e_1$ and $e_2$.  A cubic vertex on a rail changes the rail's index
and leaves the structure constant $f$; the blue arcs are the two loop
traces of each commutator, one at time $t$ and one at time $0$, which
contract the two rails of that commutator with each other and are what make
rails $1,2$ a single closed cycle and rails $3,4$ another.  The contraction
is carried out in the text.}
\label{fig:placement}
\end{figure}

%% file: references.bib
@article{wilson,
  author  = {Wilson, Kenneth G.},
  title   = {Confinement of quarks},
  journal = {Phys. Rev. D},
  volume  = {10},
  pages   = {2445--2459},
  year    = {1974},
  doi     = {10.1103/PhysRevD.10.2445}
}

@book{makeenko_methods_2002,
  author    = {Makeenko, Yuri},
  title     = {Methods of Contemporary Gauge Theory},
  series    = {Cambridge Monographs on Mathematical Physics},
  publisher = {Cambridge University Press},
  year      = {2002},
  doi       = {10.1017/CBO9780511535147},
  isbn      = {9780521809115}
}

@article{steinberg_thermalization_2019,
  author  = {Steinberg, Julia and Swingle, Brian},
  title   = {Thermalization and chaos in {QED3}},
  journal = {Phys. Rev. D},
  volume  = {99},
  number  = {7},
  pages   = {076007},
  year    = {2019},
  doi     = {10.1103/PhysRevD.99.076007},
  eprint  = {1901.04984},
  archivePrefix = {arXiv},
  primaryClass  = {cond-mat.str-el}
}

@article{witten_2dym_1991,
  author  = {Witten, Edward},
  title   = {On quantum gauge theories in two dimensions},
  journal = {Commun. Math. Phys.},
  volume  = {141},
  pages   = {153--209},
  year    = {1991},
  doi     = {10.1007/BF02100009}
}

@article{braaten_soft_1990,
  author  = {Braaten, Eric and Pisarski, Robert D.},
  title   = {Soft amplitudes in hot gauge theories: A general analysis},
  journal = {Nucl. Phys. B},
  volume  = {337},
  pages   = {569--634},
  year    = {1990},
  doi     = {10.1016/0550-3213(90)90508-B}
}

@article{berges_gaugeinvariant_2020,
  author  = {Berges, J{\"u}rgen and Boguslavski, Kirill and Mace, Mark and Pawlowski, Jan M.},
  title   = {Gauge-invariant condensation in the nonequilibrium quark-gluon plasma},
  journal = {Phys. Rev. D},
  volume  = {102},
  pages   = {034014},
  year    = {2020},
  doi     = {10.1103/PhysRevD.102.034014},
  eprint  = {1909.06147},
  archivePrefix = {arXiv},
  primaryClass  = {hep-ph}
}

@article{hashimoto_chaoswilson_2018,
  author  = {Hashimoto, Koji and Murata, Keiju and Yoshida, Kentaroh},
  title   = {Chaos of {Wilson} loop from string motion near black hole horizon},
  journal = {Phys. Rev. D},
  volume  = {98},
  pages   = {086007},
  year    = {2018},
  doi     = {10.1103/PhysRevD.98.086007},
  eprint  = {1803.06756},
  archivePrefix = {arXiv},
  primaryClass  = {hep-th}
}

@article{roberts_operator_2018,
  author  = {Roberts, Daniel A. and Stanford, Douglas and Streicher, Alexandre},
  title   = {Operator growth in the {SYK} model},
  journal = {JHEP},
  volume  = {06},
  pages   = {122},
  year    = {2018},
  doi     = {10.1007/JHEP06(2018)122},
  eprint  = {1802.02633},
  archivePrefix = {arXiv},
  primaryClass  = {hep-th}
}

@article{stanford_many-body_2016,
  author  = {Stanford, Douglas},
  title   = {Many-body chaos at weak coupling},
  journal = {JHEP},
  volume  = {10},
  pages   = {009},
  year    = {2016},
  doi     = {10.1007/JHEP10(2016)009},
  eprint  = {1512.07687},
  archivePrefix = {arXiv},
  primaryClass  = {hep-th}
}

@article{aharony_komargodski_2013,
  author  = {Aharony, Ofer and Komargodski, Zohar},
  title   = {The Effective Theory of Long Strings},
  journal = {JHEP},
  volume  = {05},
  pages   = {118},
  year    = {2013},
  doi     = {10.1007/JHEP05(2013)118},
  eprint  = {1302.6257},
  archivePrefix = {arXiv},
  primaryClass  = {hep-th}
}

@article{aharony_hagedorndeconfinement_2004,
  author  = {Aharony, Ofer and Marsano, Joseph and Minwalla, Shiraz
             and Papadodimas, Kyriakos and Van Raamsdonk, Mark},
  title   = {The {Hagedorn}-deconfinement phase transition in weakly coupled
             large {N} gauge theories},
  journal = {Adv. Theor. Math. Phys.},
  volume  = {8},
  pages   = {603--696},
  year    = {2004},
  doi     = {10.4310/ATMP.2004.v8.n4.a1},
  eprint  = {hep-th/0310285},
  archivePrefix = {arXiv},
  primaryClass  = {hep-th}
}

@article{maldacena_bound_2016,
  author  = {Maldacena, Juan and Shenker, Stephen H. and Stanford, Douglas},
  title   = {A bound on chaos},
  journal = {JHEP},
  volume  = {08},
  pages   = {106},
  year    = {2016},
  doi     = {10.1007/JHEP08(2016)106},
  eprint  = {1503.01409},
  archivePrefix = {arXiv},
  primaryClass  = {hep-th}
}

@article{shenker_stringy_2015,
  author  = {Shenker, Stephen H. and Stanford, Douglas},
  title   = {Stringy effects in scrambling},
  journal = {JHEP},
  volume  = {05},
  pages   = {132},
  year    = {2015},
  doi     = {10.1007/JHEP05(2015)132},
  eprint  = {1412.6087},
  archivePrefix = {arXiv},
  primaryClass  = {hep-th}
}

@article{gross_taylor_1993,
  author  = {Gross, David J. and Taylor, Washington},
  title   = {Two-dimensional {QCD} is a string theory},
  journal = {Nucl. Phys. B},
  volume  = {400},
  pages   = {181--210},
  year    = {1993},
  doi     = {10.1016/0550-3213(93)90403-C},
  eprint  = {hep-th/9301068},
  archivePrefix = {arXiv}
}

@article{makeenko_notes_2025,
  author  = {Makeenko, Yuri},
  title   = {Notes on the Loop Equation in Loop Space},
  year    = {2025},
  eprint  = {2508.09705},
  archivePrefix = {arXiv},
  primaryClass  = {hep-th},
  note    = {originally ITEP/NBI preprint, 1994}
}

@article{larkin_ovchinnikov_1969,
  author  = {Larkin, A. I. and Ovchinnikov, Yu. N.},
  title   = {Quasiclassical Method in the Theory of Superconductivity},
  journal = {Sov. Phys. JETP},
  volume  = {28},
  number  = {6},
  pages   = {1200--1205},
  year    = {1969}
}

@article{shenker_stanford_2014,
  author  = {Shenker, Stephen H. and Stanford, Douglas},
  title   = {Black holes and the butterfly effect},
  journal = {JHEP},
  volume  = {03},
  pages   = {067},
  year    = {2014},
  eprint  = {1306.0622},
  archivePrefix = {arXiv},
  primaryClass  = {hep-th}
}

@article{chowdhury_swingle_2017,
  author  = {Chowdhury, Debanjan and Swingle, Brian},
  title   = {Onset of many-body chaos in the {$O(N)$} model},
  journal = {Phys. Rev. D},
  volume  = {96},
  number  = {6},
  pages   = {065005},
  year    = {2017},
  eprint  = {1703.02545},
  archivePrefix = {arXiv}
}

@article{brower_pomeron_2007,
  author  = {Brower, Richard C. and Polchinski, Joseph and
             Strassler, Matthew J. and Tan, Chung-I},
  title   = {The {P}omeron and gauge/string duality},
  journal = {JHEP},
  volume  = {12},
  pages   = {005},
  year    = {2007}
}

@article{kuraev_1977,
  author  = {Kuraev, E. A. and Lipatov, L. N. and Fadin, V. S.},
  title   = {The {P}omeranchuk singularity in nonabelian gauge theories},
  journal = {Sov. Phys. JETP},
  volume  = {45},
  pages   = {199--204},
  year    = {1977}
}

@article{balitsky_1978,
  author  = {Balitsky, I. I. and Lipatov, L. N.},
  title   = {The {P}omeranchuk singularity in quantum chromodynamics},
  journal = {Sov. J. Nucl. Phys.},
  volume  = {28},
  pages   = {822--829},
  year    = {1978}
}

@book{lebellac_thermal_1996,
  author    = {Le Bellac, Michel},
  title     = {Thermal Field Theory},
  publisher = {Cambridge University Press},
  year      = {1996}
}

@article{blaizot_iancu_2002,
  author  = {Blaizot, Jean-Paul and Iancu, Edmond},
  title   = {The quark-gluon plasma: collective dynamics and hard thermal loops},
  journal = {Phys. Rept.},
  volume  = {359},
  pages   = {355--528},
  year    = {2002},
  eprint  = {hep-ph/0101103},
  archivePrefix = {arXiv}
}

@article{mezei_stanford_2017,
  author  = {Mezei, M{\'a}rk and Stanford, Douglas},
  title   = {On entanglement spreading in chaotic systems},
  journal = {JHEP},
  volume  = {05},
  pages   = {065},
  year    = {2017},
  eprint  = {1608.05101},
  archivePrefix = {arXiv},
  primaryClass  = {hep-th}
}

@article{grozdanov_kinetic_2019,
  author  = {Grozdanov, Sa\v{s}o and Schalm, Koenraad and Scopelliti, Vincenzo},
  title   = {Kinetic theory for classical and quantum many-body chaos},
  journal = {Phys. Rev. E},
  volume  = {99},
  pages   = {012206},
  year    = {2019},
  eprint  = {1804.09182},
  archivePrefix = {arXiv},
}

@article{pisarski_moving_1993,
  author  = {Pisarski, Robert D.},
  title   = {Damping rates for moving particles in hot {QCD}},
  journal = {Phys. Rev. D},
  volume  = {47},
  pages   = {5589--5600},
  year    = {1993},
  eprint  = {hep-ph/9302242},
  archivePrefix = {arXiv}
}

@article{flechsig_rebhan_schulz_1995,
  author  = {Flechsig, F. and Rebhan, A. K. and Schulz, H.},
  title   = {Infrared sensitivity of screening and damping in a quark-gluon plasma},
  journal = {Phys. Rev. D},
  volume  = {52},
  pages   = {2994--3002},
  year    = {1995},
  eprint  = {hep-ph/9502324},
  archivePrefix = {arXiv}
}

@article{linde_infrared_1980,
  author  = {Linde, A. D.},
  title   = {Infrared problem in the thermodynamics of the {Yang-Mills} gas},
  journal = {Phys. Lett. B},
  volume  = {96},
  pages   = {289--292},
  year    = {1980}
}

@article{laine_realtime_2007,
  author  = {Laine, M. and Philipsen, O. and Romatschke, P. and Tassler, M.},
  title   = {Real-time static potential in hot {QCD}},
  journal = {JHEP},
  volume  = {03},
  pages   = {054},
  year    = {2007},
  eprint  = {hep-ph/0611300},
  archivePrefix = {arXiv}
}
